\documentclass[10pt,journal,compsoc,twoside]{IEEEtran}
   
\ifCLASSINFOpdf
  \usepackage[pdftex]{graphicx}
\else
  \usepackage[dvips]{graphicx}
\fi
 
\usepackage{cite}
\usepackage{amsmath,amssymb,amsfonts}
\usepackage{algorithmic}
\usepackage{graphicx}
\usepackage{textcomp}

\usepackage{pifont}
\usepackage{algorithm}
\usepackage{array}
\usepackage{url}
\usepackage[backref]{hyperref} 
\hypersetup{
colorlinks=true,
linkcolor=black,
citecolor=black,
urlcolor=black
} 
\PassOptionsToPackage{dvipsnames,table,xcdraw}{xcolor}
\newif\ifdraft
\draftfalse 

\ifdraft
    \usepackage[authormarkup=none, nocomment]{changes}
\else
    \usepackage[final, nocomment]{changes}
\fi
\definechangesauthor[name={Addition}, color=black]{ADD}
\definechangesauthor[name={Replacement}, color=black]{REP}
\definechangesauthor[name={Deletion}, color=Red]{DEL}
\definechangesauthor[name={Final}, color=Red]{FIN}

\usepackage{tcolorbox}
\usepackage{multirow,tabularx} 
\usepackage{colortbl}
\usepackage{footmisc} 
\usepackage{wrapfig}
\usepackage{tablefootnote}
\usepackage{siunitx} 
\usepackage{booktabs}
\usepackage{makecell}
\usepackage{siunitx}
\usepackage{threeparttable}

\ifCLASSOPTIONcompsoc
  \usepackage[caption=false,font=footnotesize,labelfont=sf,textfont=sf]{subfig}
\else
  \usepackage[caption=false,font=footnotesize]{subfig}
\fi

\begin{document}

\title{
  BASE: Burst-Adaptive Autoscaling via Stacked Ensembles for SLO Assurance and Cost Efficiency
}

\author{Chunyang Meng, Haogang Tong, Tianyang Wu, Maolin Pan, Yang Yu{*}, and Yi Jiang{*}
\IEEEcompsocitemizethanks{\IEEEcompsocthanksitem C. Meng and Y. Jiang
are with the School of Computer Science and Technology, Chongqing University of Posts and Telecommunications, Chongqing, China.
\protect\\E-mail: \{mengcy, jiangyi\}@cqupt.edu.cn.
\IEEEcompsocthanksitem H. Tong and T. Wu are with the School of Computer Science and Engineering, Sun Yat-sen University, Guangzhou, China.
\protect\\E-mail: \{tonghg, wuty26\}@mail2.sysu.edu.cn.
\IEEEcompsocthanksitem M. Pan and Y. Yu are with the School of Software Engineering, Sun Yat-sen University, Zhuhai, China.\protect\\E-mail: \{panml, yuy\}@mail.sysu.edu.cn.
}
\thanks{\\Manuscript received Aug. 5, 2025; revised Dec. 27, 2025; accepted Feb. 22, 2026. (Corresponding author: Yang Yu and Yi Jiang.)}
}

\markboth{IEEE Transactions on Services Computing
}%
{Meng et al.: BASE: Burst-Adaptive Autoscaling via Stacked Ensembles for SLO Assurance and Cost Efficiency}

\IEEEtitleabstractindextext{%
\begin{abstract}

Autoscaling is a technology that automatically scales resources for applications without human intervention to ensure runtime Quality of Service (QoS) while reducing costs. However, user-facing cloud applications serve dynamic workloads that often exhibit variability and contain bursts, posing challenges to autoscaling in maintaining QoS within Service-Level Objectives (SLOs). Conservative strategies risk over-provisioning, while aggressive ones may cause SLO violations, making it more challenging to design effective autoscaling.  
This paper introduces BASE, a burst-adaptive autoscaling framework that leverages a stacked ensemble of machine learning models to mitigate SLO violations and reduce costs for containerized services and applications operating under time-varying workloads.
BASE incorporates a novel prediction-based burst detection mechanism that distinguishes between predictable workload spikes and actual uncertain bursts. When bursts are detected, BASE appropriately overestimates them and allocates resources accordingly to address the rapid growth in resource demand. On the other hand, BASE employs reinforcement learning to rectify potential inaccuracies in resource estimation, enabling more precise resource allocation during non-burst periods.  
Experiments across ten real-world workloads demonstrate BASE's effectiveness, achieving a significant reduction in SLO violations with lower resource costs compared to other prominent methods.  

   \end{abstract}

}
\maketitle

\IEEEdisplaynontitleabstractindextext
\IEEEpeerreviewmaketitle

\IEEEraisesectionheading{\section{Introduction}\label{sec_introduction}}

\IEEEPARstart{C}{loud} computing is gaining traction, as evidenced by the widespread use of cloud-based software services or applications~\cite{10433234}. 
To date, three main service models have fostered cloud adoption, namely Software, Platform, and Infrastructure as a Service (SaaS, PaaS, and IaaS)~\cite{buyya2018manifesto}.
Among all these models, one of the pronounced benefits of the cloud is elasticity, which allows subscribers to acquire or release resources provided by their applications on demand as workloads change~\cite{10419899}, and to pay only for what has been used~\cite{9744560}.
Therefore, cloud computing reduces costs for subscribers, from small businesses to large enterprises, and improves resource utilization, which has often been low in the past~\cite{li2022serverless}.

\begin{figure}[htbp]
  \hspace{-1em}
  \includegraphics[width=1.05\linewidth]{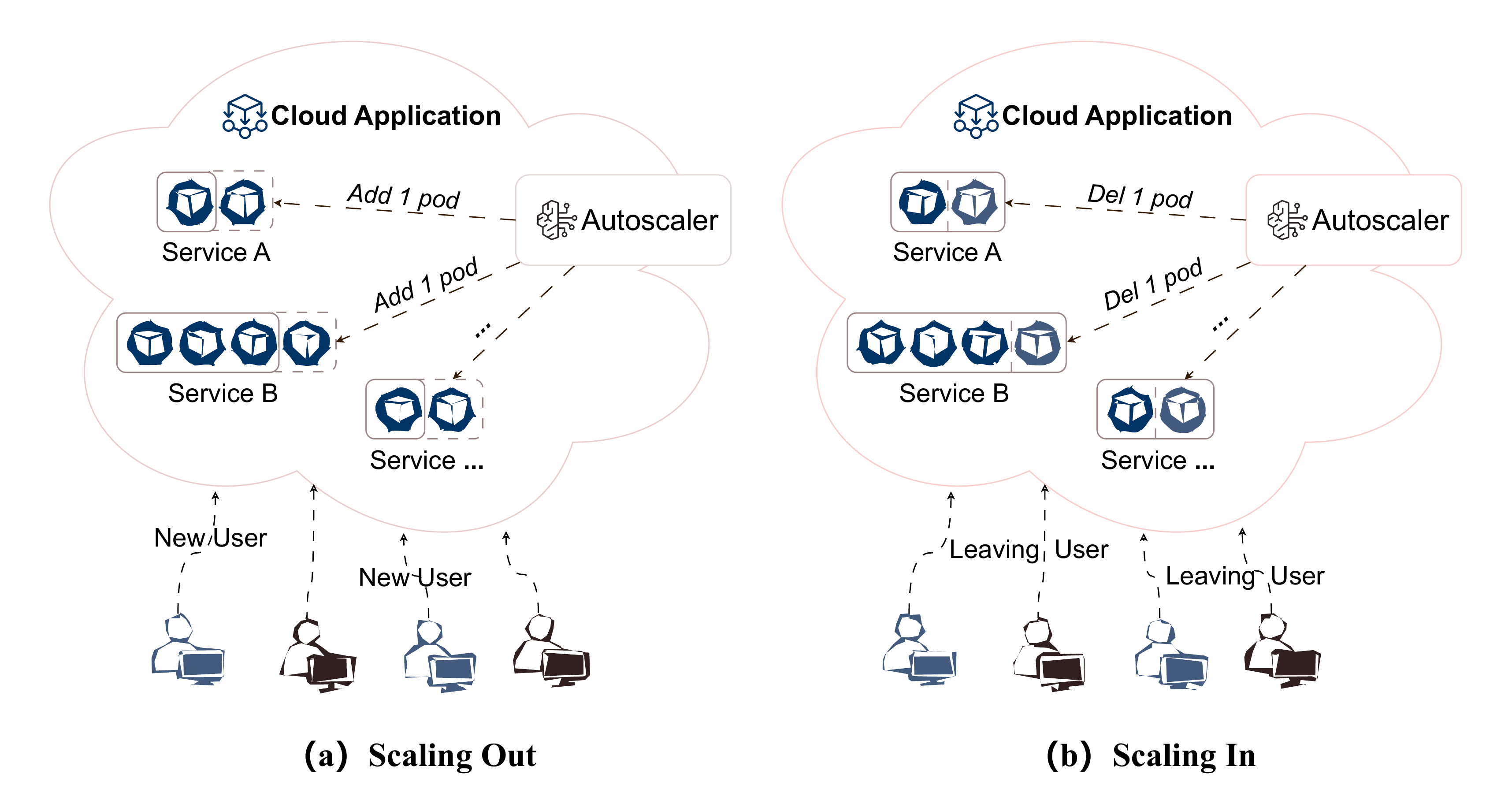}
  \caption{Typical autoscaling scenarios—right sizing of resources. (a) Increased requests lead to congestion, prompting the autoscaler to scale out (adding pods).
  (b) Decreased requests trigger deprovisioning actions, such as scaling in (removing pods).
  }
  \label{fig_intro_autoscaling}
\end{figure}
Despite these advantages, on-demand resource management still faces challenges. Due to rapid changes and fierce competition in the market environment, cloud applications experience fluctuations in user requests~\cite{alibabacloud}.
When such fluctuations occur, adjusting the amount of resources as early as possible is necessary.
Otherwise, the service performance may be compromised when the request workload increases, or resources may be wasted when the request workload decreases.
Manual synchronization of resource provisioning is complex and lacks real-time assurance~\cite{TARI2024100650}, 
\added[id=ADD]{a complexity further amplified in modern cloud-native applications structured as microservices, whose intricate dependency graphs allow localized workload bursts to propagate downstream, triggering cascading SLO violations that defy manual intervention.} 
\replaced[id=REP]{The desire to overcome these issues motivates the adoption of autoscaling, as shown in Fig.~\ref{fig_intro_autoscaling}, which autonomously and dynamically provisions and de-provisions resources to accommodate fluctuating request workloads without human intervention, ensuring that resource consumption is minimized while QoS is maintained~\cite{qu2018auto}.}

Over the years, many efforts have been devoted to workload autoscaling, primarily using rule-based reactive methods~\cite{yu2020microscaler,ding2021copa,tadakamalla2020autonomic}, predictive proactive methods~\cite{imdoukh2020machine,golshani2021proactive,ramperez2021flas,qian2022robustscaler,chen2016self}, and hybrid methods~\cite{bauer2018chameleon,chen2022resource,schuler2020ai,yan2021hansel}, aiming to achieve effective and dependable runtime scaling.   
However, we identify three main limitations of these methods:
\begin{enumerate}
  \item First, most of them are oblivious to workload bursts and lack targeted solutions, making them vulnerable to bursts.
  \item A few works focus on bursts~\cite{abdullah2020burst,trihinas2017improving}, but the burst detection algorithms in these works use statistical (e.g., moving average~\cite{vlachos2004identifying} and entropy~\cite{ali2014measuring}) approaches that detect bursts by quantifying workload fluctuations. As a result, they may mislabel predictable workload peaks as bursts, thus affecting overall performance.
  \item Relying solely on direct resource prediction models for resource estimation suffers from a scarcity of labeled data and unavoidable prediction errors, leading to a lack of robustness and suboptimal scaling decisions.

\end{enumerate}

\textbf{The BASE Framework.} 
To this end, we present BASE, a real-time burst-adaptive autoscaling framework for containerized cloud services and applications. It aims to dynamically allocate resources by accounting for workload bursts and leveraging a stacked ensemble of machine learning (ML) models, thereby enhancing SLO assurance and cost efficiency under time-varying workloads. BASE operates through four key components: \emph{Workload Prediction}, \emph{Burst Detection and Handling}, \emph{Resource Estimation}, and \emph{Estimation Enhancement}
The first component, \emph{Workload Prediction}, forecasts future workload ranges based on a Transformer model and historical data. The \emph{Burst Detection and Handling} module identifies bursts by comparing predicted and observed workloads, allowing BASE to distinguish predictable workload spikes from genuinely uncertain bursts. Correspondingly, BASE applies different resource allocation strategies depending on whether a burst occurs.
For burst scenarios, BASE employs an Autoregressive (AR) model to capture short-term trends and derive an overestimation via Bootstrapping, and then integrates Support Vector Regression (SVR) models for resource estimation. For non-burst scenarios, BASE further employs deep reinforcement learning (DRL) to refine resource allocation in real time.

\replaced[id=ADD]{Crucially, BASE adapts to microservice architectures by leveraging the Transformer's joint modeling of multivariate to implicitly learn workload propagation dynamics between coupled services directly from historical data. This data-driven approach eliminates the need for maintaining static dependency graphs and, when combined with the rapid response of the \emph{Burst Detection and Handling} component, effectively mitigates cascading SLO violations.}{}
\replaced[id=REP]{Furthermore, by integrating workload forecasting and performance modeling to indirectly estimate resource needs—rather than relying on supervised labels—BASE eliminates the need for costly expert annotations.
Despite utilizing a stacked ensemble, the system maintains low runtime overhead due to sequential execution, with the computational cost dominated primarily by the most complex model (i.e., Transformer) in the pipeline.}{}

\textbf{Contributions.} 
To the best of our knowledge, this is the first work to enable burst-adaptive autoscaling for containerized cloud services and applications through robust resource estimation via a stacked ensemble of ML models, eliminating the need for labeled data while improving QoS assurance and cost efficiency.
In summary, our main contributions are as follows:

\begin{enumerate}  \item \textbf{\emph{Burst Detection and Handling}}: We propose a novel prediction-based mechanism for burst detection, which treats workloads that do not conform to predictions as bursts.
  Moreover, we utilize the AR model and \emph{Bootstrapping} to obtain an appropriate overestimation of burst intensity, providing valuable references for resource planning.
  \item \textbf{\emph{Estimation Enhancement}}: We propose a method to enhance resource estimation. It improves the accuracy and robustness of resource allocation by dynamically correcting potentially inaccurate estimates using a DRL agent and real-time telemetry data.

  \item \textbf{\emph{Burst-Adaptive Autoscaling}}: We propose a robust burst-adaptive autoscaling framework, named BASE, which performs targeted resource planning depending on whether the workload is bursty and allocates resources accordingly, optimizing SLO assurance and cost efficiency under fluctuating workloads.

  \item \textbf{\emph{Implementation and Evaluation}}: We implemented the BASE prototype and validated it on five benchmark applications and ten real-world workloads in a Kubernetes-based environment. The experimental results show that BASE is a promising approach.
\end{enumerate}

\section{Background and Motivation}
\subsection{Background: Workload Bursts}

The large-scale social behaviour inherent in cloud-native systems leads to workload bursts in user requests~\cite{calzarossa2016workload}. Such bursts can originate from flash sales, viral content, or unpredictable user behavior, and can rapidly saturate available resources, thereby triggering SLO violations, increased response times, or even cascading system failures~\cite{yin2014system}.

\begin{table}[h]
\centering
\caption{Current Measures of Workload Bursts}
\label{tab:burst_measures}
\renewcommand{\arraystretch}{1.3}
\begin{tabular}{|c|c|}
\hline
\textbf{Measure} & \textbf{Details} \\ \hline
\makecell{Auto-Correlation\\Function, ACF} & $ACF(t, b) = \frac{E[(X_t - \mu_t)(X_b - \mu_b)]}{\sigma_t \sigma_b}$ \\ \hline
Index of Dispersion, $I$ & $I = SCV(1 + 2\sum_{i=1}^{\infty} \rho_i)$ \\ \hline
Hurst Parameter, $H$ & $VAR[X^{(m)}] \sim km^{2H-2}$ as $m \to \infty$ \\ \hline
\makecell{Deviation from\\Moving Average, $D(X)$} & $D(X) = avg(X) + \eta \times std(X)$ \\ \hline
Shannon Entropy, $H_S(X)$ & $H_S(X) = -\sum_{i=1}^{n} P(X_i) \log P(X_i)$ \\ \hline
\end{tabular}
\end{table}

Although burstiness is widely recognized as a critical factor affecting QoS, there is no universally accepted formal definition of bursts in workload traces. As a result, burst detection in the literature typically relies on analyzing statistical properties of the workload time series rather than applying a fixed threshold or structural model.
Table~\ref{tab:burst_measures} summarizes commonly used statistical measures for quantifying burstiness in workload time series $X_t \ (t = 1,2,3,\dots)$. 
In this table, $E$ denotes the expected value of $X$; $\theta_t$ and $\theta_b$ are the standard deviations at lags $t$ and $b$; $SCV$ represents the squared coefficient of variation; $\rho$ is the autocorrelation coefficient; $k$ is a constant; and $m$ is the sample size. The symbol $H$ refers to the Hurst exponent (bounded within $[0.5,1]$); $\eta$ is a generic coefficient; and $P(X_i)$ indicates the probability that $X$ equals the value $X_i$.

The Auto-Correlation Function (ACF) is frequently applied to reveal repeating patterns in the series. The Index of Dispersion, obtained by normalizing the variance with the mean, highlights the degree of over-dispersion. The Hurst exponent gauges long-range temporal dependence—larger values imply stronger persistence and, hence, more pronounced burstiness. The Deviation from a Moving Average quantifies how far a specific observation strays from its smoothed baseline, offering a straightforward local signal of transient bursts. Finally, Shannon Entropy measures the uncertainty within the workload distribution and excels at pinpointing abrupt departures from routine behavior. Collectively, these measures furnish complementary perspectives for analyzing burst characteristics in cloud-workload traces.

\begin{figure}[htbp]
  \centering
  \includegraphics[width=\linewidth]{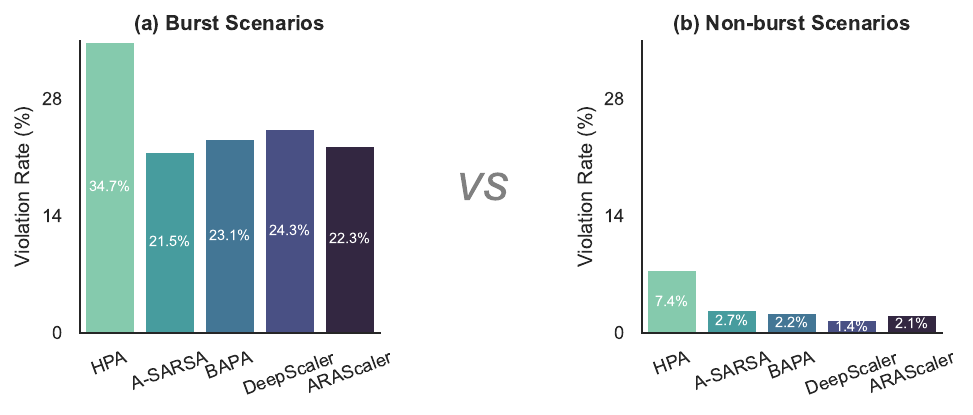}
  \caption{SLO Violation Comparison By Scenario}
  \label{fig_motivation_burst}
\end{figure}

\subsection{Motivation}
\label{sec_motivation}

\textbf{Motivation \ding{172}: Bursty workloads pose a severe threat to effective autoscaling.} 
Maintaining consistent QoS under bursty workloads still remains a critical challenge in modern cloud-native systems. Although autoscaling mechanisms are widely adopted to dynamically adjust resources in response to workload changes, their effectiveness and accuracy under bursty workloads are often insufficient, leading to elevated SLO violations and poor user experience.
Traditional reactive autoscalers, such as the kubernetes Horizontal Pod Autoscaler (HPA)~\cite{k8sHPA}, suffer from inherent scaling lag in bursty scenarios due to their reliance on delayed metric feedback. 
In contrast, predictive autoscalers like DeepScaler~\cite{DeepScaler} aim to anticipate future demand, but their accuracy deteriorates sharply in the presence of bursts with high variability and uncertainty. 
As a result, both reactive and predictive approaches tend to exhibit suboptimal scaling behavior under bursts—either responding too late or acting on unreliable predictions.
We evaluated the impact of bursts on four representative baseline autoscaling methods using 10 real-world workloads (see Section 4.1 for details of the experimental settings). As shown in Fig.~\ref{fig_motivation_burst}, all baselines maintained low SLO violation rates (1.4\%--7\%) under non-bursty conditions. However, in bursty scenarios, the violation rates increased drastically---by factors ranging from 5$\times$ to 17$\times$---revealing severe performance degradation.
The pronounced performance gap highlights the inadequacy of existing mechanisms in handling bursts, and motivates the need for autoscaling approaches that incorporate burst detection and handling capabilities to maintain QoS under bursty workloads.

\begin{figure}[htbp]
  \includegraphics[width=\linewidth]{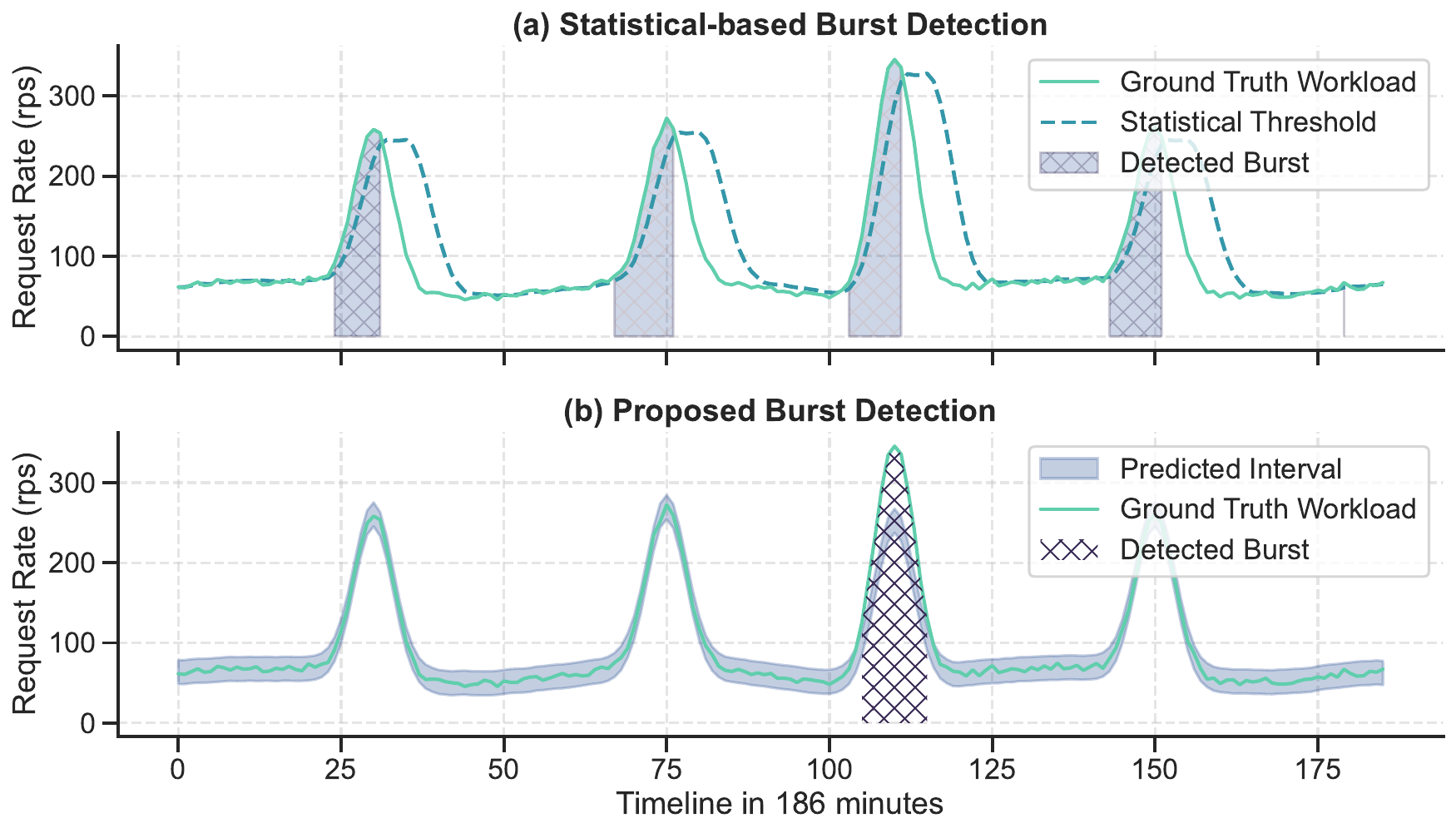}
  \caption{An example illustrates the comparison. The proposal will not identify periodic workload spikes as bursts. Other statistical methods primarily differ in statistical threshold $ST_w$. Here, the $ST_w=avg(X)+0.8\times std(X)$ presented in \cite{vlachos2004identifying}.}  
  \label{fig_comparison_burst_detection}
\end{figure}

\begin{figure*}[htbp]
  \centering
  \includegraphics[width=0.95\linewidth]{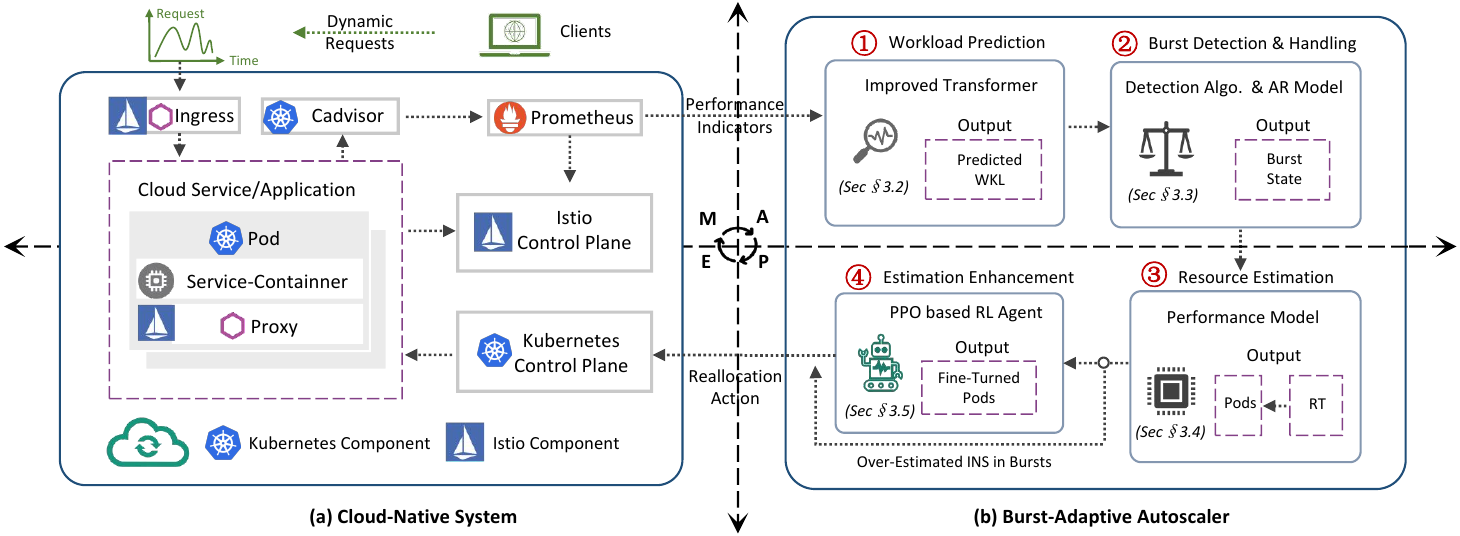}
  \caption{Overview of the BASE shown as a MAPE loop}
  \label{fig:overview}
\end{figure*}

\textbf{Motivation \ding{173}: Predictable workload peaks should not be mistaken for bursts.} 
As discussed in Motivation 1, workload burstiness is a key factor affecting both the efficiency of resource decisions and the quality of service. Since bursts often lead to sharp increases in request volume, the common strategy is to provision excess resources to enhance the system's responsiveness. However, this "redundancy-for-stability" approach introduces another issue: when normal workload fluctuations are misidentified as bursts, it can result in resource waste and increased operational costs.
Traditional burst detection methods typically rely on statistical indicators—such as ACF, index of dispersion, and deviation from moving average—to quantify workload volatility, and use fixed thresholds to determine whether a burst has occurred~\cite{abdullah2020burst}. However, such methods fail to distinguish between genuine bursts and predictable peaks. With the rapid advancement of time-series forecasting techniques~\cite{yan2021hansel,masdari2020survey}, many previously unpredictable "bursty" workloads can now be anticipated with reasonable accuracy using deep learning models such as Informer~\cite{zhou2021informer} and TimeMixer~\cite{wang2024timemixer}. This suggests that some workload peaks are not truly bursty events, but rather predictable and manageable trends that do not require reactive resource over-provisioning.


Under such circumstances, continuing to use volatility-based statistical detection may result in misclassifying predictable peaks as bursts. This not only triggers unnecessary resource scaling but also disrupts the steady-state scheduling behavior of autoscalers, reducing overall resource efficiency. This highlights the need for burst detection strategies that incorporate workload predictability to more accurately differentiate between true bursts and expected demand peaks.
Fig.~\ref{fig_comparison_burst_detection} presents a comparison between traditional statistical burst detection methods and our prediction-based approach. While conventional methods classify all workload peaks as bursts, our method identifies only those deviations that fall outside the forecast confidence interval, significantly improving the accuracy of burst detection and resource efficiency.

\textbf{Motivation \ding{174}: Label Scarcity and Forecast Error in Predictive Autoscaling.}
Predictive autoscaling aims to proactively allocate resources by forecasting future resource demands, thereby avoiding the scaling lag inherent in reactive approaches. However, its practical effectiveness is constrained by two fundamental issues~\cite{qian2022robustscaler}. First, training accurate predictive models typically requires a large amount of high-quality labeled data (e.g., optimal resource configurations satisfying QoS constraints), which is expensive and scarce in real-world environments. Moreover, the diversity of microservice scenarios and user behaviors severely limits the generalization capability of such models. Second, even with high-quality labeled data, prediction methods cannot eliminate forecast errors caused by the inherent uncertainty of cloud environments, especially under bursty workloads, where prediction accuracy tends to degrade significantly.  As illustrated in Fig.~\ref{fig_motivation_burst}, inaccurate predictions by forecasting methods (e.g., A-SARSA) under bursty conditions directly lead to significantly increased SLO violations. This underscores the limitations of existing predictive models and suggests the need for a new autoscaling mechanism that reduces reliance on costly labeled data while remaining robust to prediction inaccuracies.

\section{The BASE Framework}

\subsection{System Overview}

The overview of the BASE framework is presented in Fig.~\ref{fig:overview},
 llowing a typical MAPE loop: Monitoring (M), Analysis (A), Planning (P), and Execution (E). We delegate monitoring and execution to cloud-hosted third-party software or call the cloud provider's API. This paper focuses on the analysis and planning phases, with Algorithm~\ref{alg:autoscaling} describing the workflow of the Burst-adaptive Autoscaler.

\begin{itemize}
  \item \textbf{Monitoring.} In monitoring phase, BASE needs to systematically monitor certain performance metrics to determine the necessity and methodology for scaling operations (shown at the top of Fig.~\ref{fig:overview}(a) and described in Line 5 of Algorithm~\ref{alg:autoscaling}).
  Specifically, BASE uses the cAdvisor, integrated into the Kubernetes, to monitor the applications and collect metrics at the container layer, including CPU utilization, memory usage, etc. In addition, it uses Istio to generate detailed telemetry for all service communications and collect metrics at the application layer (e.g., request rate and average response time). The collected metrics are stored in Prometheus, a time series database, for querying.

  \item \textbf{Analysis.} In analysis phase, the collected performance metrics undergo further processing to unveil latent information crucial for determining scaling operations. BASE first uses the monitored workloads and the improved Transformer with long-term patterns capture capabilities to predict the range of future workloads (marked as \textcolor{red}{\textbf{\large\textcircled{\small 1}}}, described in Line 6, and details in §\ref{sec:method_wkl_predict}).
  Based on the prediction results, real-time comparison between the predicted and actual workloads is then conducted to detect bursts
  (marked as \textcolor{red}{\textbf{\large\textcircled{\small 2}}}, described in Line 7, and details in §\ref{sec:method_burst_detection_and_handling}).
  Once the burst is detected, BASE utilizes an AR model with short-term trend capture capabilities to generate an appropriate overestimation via Bootstrapping method
  (marked as \textcolor{red}{\textbf{\large\textcircled{\small 2}}}, described in Lines 8-10, and details in §\ref{sec:method_burst_detection_and_handling}).

  \item \textbf{Planning.} The planning phase estimates the total amount of resources that should be allocated in the next scaling action. Based on whether a burst occurs or not, BASE adopts different resource estimation methods. 
  For bursts, BASE utilizes performance models, the overestimated burst workloads, and performance constraints (i.e., SLO) to generate a relatively generous amount of resources
  (marked as \textcolor{red}{\textbf{\large\textcircled{\small 3}}}, described in Line 11, and details in §\ref{sec:method_resource_estimation}).
  For non-bursts, BASE employs the performance model for initial resource estimation and then jointly deep reinforcement learning to correct possible inaccuracy estimation dynamically
  (marked as \textcolor{red}{\textbf{\large\textcircled{\small 4}}}, described in Lines 12-14, and details in §\ref{sec:method_estimation_enhance}). 

  \item \textbf{Execution.} The execution phase provisions or deprovisions resources based on the scaling plan
  (shown at the bottom of Fig.~\ref{fig:overview}(a) and described in Line~15).
  It is straightforward, and we implemented it by calling the Kubernetes API~\cite{k8sapi}.
\end{itemize}

\begin{algorithm}[htbp]
  \caption{Pseudocode of the BASE}
  \label{alg:autoscaling}
  \begin{algorithmic}[1] 
      \STATE Initialize workload predictor $Q$ with trained weights $\sigma$
      \STATE Initialize performance model $P$ with trained weights $\omega$
      \STATE Initialize DRL policy $\pi_{\theta}$ with trained parameters $\theta$ 
      \WHILE{not done}
          \STATE Observe $S_t=\{ins_t, cpu_t, mem_t, wkl_t\}$ and time-series workload $X_t=\{wkl_{t-M},...,wkl_{t-1}, wkl_t\}$
          \STATE Predict workload $\hat{Y}_{t}=Q_{\sigma}(X_{t})$ and set $wkl_{t}=\hat{Y}_{t}[1][0]$
          \STATE Detect burst state $B^t$ by Algo.~\ref{alg:burst_detection} using $\hat{Y}_{t}$
          \IF{$B^t$ is burst}
          \STATE Overestimate bursty workload by AR model, \\and set to $wkl_{t}$
          \ENDIF
          \STATE Estimate resource $ins_t$ by:\\
          $\arg\mathop{\max}\limits_{INS}P_{\omega}(RT |$\emph{INS}$, wkl_{t}, SVC),\ s.t.\  RT<\lambda_{RT}$
          \IF{$B^t$ is not burst}
          \STATE Enhance resource estimation by PPO policy:\\
          \ \ \ \ \ \ \ $ins_t=A_t=\pi_\theta(a|s=S_t)$
          \ENDIF
          \STATE Allocate resource by $ins_t$
          \STATE Wait until next time step, and set $t=t+1$
      \ENDWHILE
  \end{algorithmic}
\end{algorithm}\subsection{Workload Prediction}\label{sec:method_wkl_predict}

The workload predictor is designed to forecast the ranges of user request workloads for multiple consecutive timesteps in the future. We employ an improved Transformer structure~\cite{vaswani2017attention} to analyze and learn from the request workload of the running applications. Compared to other time series prediction models, the transformer architecture naturally supports multiple sequence predictions, allowing it to predict multiple workload time series simultaneously (e.g., workloads of microservice applications composed of multiple services). 
In addition, by introducing the sparse attention mechanism proposed by Zhou et al.~\cite{zhou2021informer}, the efficiency of real-time workload prediction is guaranteed.

\textbf{Workload Representation.} Concretely, we consider an application composed of $N$ microservices, each associated with a historical workload time series observed in the past $L_x$ time slices. Each time slice contains a timestamp string and a scalar request rate. To capture workload patterns, we embed three kinds of context: (1) \emph{Positional embedding}: We apply a sinusoidal encoding with alternating sine and cosine functions to represent the relative positions of time steps; (2) \emph{Temporal embedding}: Calendar-based features (e.g., hour-of-day) are normalized to encode periodic patterns commonly observed in cloud workloads; and (3) \emph{Value embedding}: Raw workload values are passed through a 1-D convolutional layer to obtain dense representations that capture short-term local trends. The final input sequence $\mathcal{X}^{t}_{\operatorname{feed}}$ is derived by the element-wise sum of these three embeddings.

\textbf{Encoder-Decoder.} 
The feeding matrix $\mathbf{X}^{t}_{\operatorname{en}}$ is then passed through the encoder and decoder modules to generate the workload forecast. 
The encoder contains a three-layer stack and a one-layer stack (with a quarter of the input), and each stack propagates from the $j$-th layer to the $(j+1)$-th layer as:
\begin{equation}
  \mathbf{X}^t_{j+1} = \operatorname{MaxPool}(\operatorname{ELU}(\operatorname{Conv1d}([\mathbf{X}^t_j]_{\operatorname{AB}})))
  \label{eq_encoder}
\end{equation}
where $[\cdot]_{\operatorname{AB}}$ represents the \emph{Multi-head ProbSparse Self-attention} layer~\cite{zhou2021informer}. $\operatorname{Conv1d}(\cdot)$ performs 1-D convolution (kernel width=3) along the time dimension with the $\operatorname{ELU}(\cdot)$ activation function. Then, a max-pooling layer $\operatorname{MaxPool}(\cdot)$ with stride=2 downsamples $\mathbf{X}^t$ to half its length after each stacked layer. The final hidden representation of the encoder is the concatenation of the outputs from all stacks.

Then, the decoder module decodes $\mathcal{H}^t$ into an output representation $\hat{\mathcal{Y}}^t$. 
It is composed of a stack of two identical multi-head attention layers. ServiceID-wise masked multi-head attention is applied in the \emph{ProbSparse} self-attention computation by setting masked dot products to $-\infty$. 
Finally, we apply a shared linear projection \(\mathrm{Proj}_{\mathrm{out}}\!:\mathbb{R}^{d_h}\to\mathbb{R}^{|Q|}\) to each vector in  
\(\mathbf{Y}_{L_d}^t\) and reshape to  
\(\hat{\mathcal{Y}}^t\in\mathbb{R}^{L_y\times N\times |Q|}\), where  
\(\hat{y}^{t}_{i,n,q}\) is the predicted \(q\)-quantile for service \(n\) at step \(i\).  
This produces simultaneous forecasts for all \(N\) sequences over the horizon \(L_y\). 

We choose the quantile loss function on result w.r.t the workload forecasting, as it can compute the loss between ground truth $\mathcal{Y}^t=\{y^{t}_1,...,y^{t}_{L_y}\}$ and the prediction $\hat{\mathcal{Y}}^t$ in given quantiles $Q$:

\begin{equation}
  \mathcal{L}_q(y,\hat{y}) =q\times \max(y-\hat{y},0) + (1-q)\times \max(\hat{y} - y,0)
  \label{eq_q_loss}
\end{equation}
\begin{equation}
  \mathcal{L}_Q(\mathcal{Y}^t,\hat{\mathcal{Y}}^t) =\frac{1}{N|Q|} \sum_{n=1}^{N}\sum_{i=1}^{L_y}\sum_{q\in Q} \mathcal{L}_q(y^{t}_{i,n},\hat{y}_{i,n,q}^{t})
  \label{eq:predictor_loss}
\end{equation}
In our experiments, the joint loss of the prediction interval's upper, median, and lower bounds can be calculated by setting $Q=\{0.1,0.5,0.9\}$.
Moreover, the parameters are optimized by \emph{Stochastic Gradient Descent} (via \emph{Adam}) that propagated back the loss from the decoder's outputs across the entire model.

\begin{figure}[htbp]
	\centering
	\includegraphics[width=\linewidth]{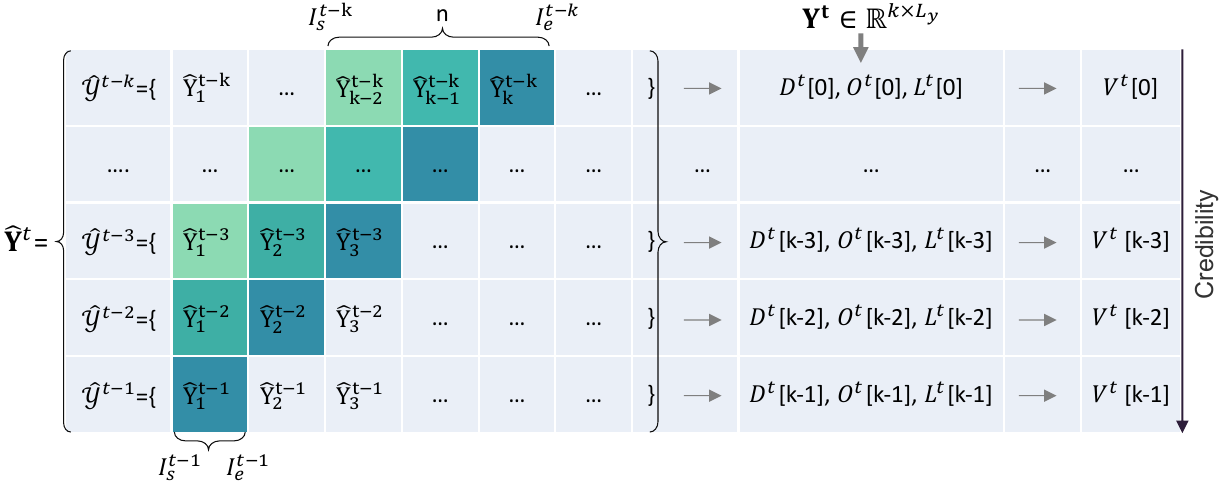}
	\caption{Calculation process of $V^t$}
	\label{fig_burst_identifie}
\end{figure}

\subsection{Burst Detection and Handling}\label{sec:method_burst_detection_and_handling}

The burst detector and handler component is designed to detect bursty workloads and provide an effective way to handle them accordingly.

\textbf{Burst Detection.} 
Although true bursts are inherently uncertain, a significant portion of workload variations exhibits clear and predictable patterns. This observation motivates a key insight: while bursts may not be directly predicted, they can be indirectly identified by learning the normal predictable patterns and comparing them against the actual runtime workloads. Building on this idea, the proposed burst detector takes as input the predictions over the past $k$ time steps for each service, denoted as $\hat{\mathbf{Y}}^t_n=\{\hat{\mathcal{Y}}^{j}_n=\{\hat{Y}^j_{1,n},...,\hat{Y}^j_{L_y,n}
\}|j\in[t-k,t-1]\}\in \mathbb{R}^{k\times L_y \times |Q|}$, and compares them with the ground truths $\mathbf{Y}^t_n=\{\mathcal{Y}^{i}_n=\{y^{j}_1,...,y^{j}_{L_y}\}|j\in[t-k,t-1]\}\in \mathbb{R}^{k\times L_y}$ to determine whether the workload at time $t$ is bursty. To this end, we designed $D^t\in \mathbb{R}^k$ and $O^t \in \mathbb{R}^k$ to quantify the deviation distances and the number of deviations for each $\mathcal{Y}^j\mbox{-}\hat{\mathcal{Y}}^j$ pair in the last $k$ time steps of each service, respectively:

\begin{equation}
  \label{eq_D}
  D^t = \{\frac{1}{I_e-I_s+1}\sum_{i=I_s}^{I_e} \operatorname{d}(\hat{Y}^j_i, y^{j}_i)\ |\ j\in [t-k,t-1]\} 
\end{equation}
\begin{equation}
  \label{eq_O}
  O^t = \{\sum_{i=I_s}^{I_e}(\operatorname{d}(\hat{Y}^j_i, y^j_i)>0)\ |\ j\in [t-k,t-1]\}
\end{equation}
where $I_s^j$ and $I_e^j$, defined in Eq.~\eqref{eq_index}, denote the starting and ending indexes of the $y^{j}_{\cdot}\mbox{-}\hat{Y}^j_{\cdot}$ pair that need to be calculated at time $j$, respectively. 
$\operatorname{d}(\cdot,\cdot)$$\ge$0 is defined in Eq.~\eqref{eq_distance}. 
When the truth is within the prediction interval, then $\operatorname{d}(\cdot,\cdot)$=0. Conversely, $\operatorname{d}(\cdot,\cdot)$$>$0 is called an outlier, and its value reflects the relative distance between the outlier and the prediction interval. 
\begin{equation}
  I_e^j=t-j;\ 
  I_s^j=\left\{
  \begin{aligned}
   & 1, &\operatorname{if} I_e^j \le n \\
   & I_e-n+1, &\operatorname{else}
  \end{aligned}
  \right.
  ;\ j\in [t-k,t-1]
  \label{eq_index}
\end{equation}
\begin{equation}
  \operatorname{d}(\hat{Y}^t_i, y^{t}_{i}) = \operatorname{ReLU}(y^{t}_{i} - \hat{y}^{t}_{i,0.9})/\hat{y}^{t}_{i,0.9} + \operatorname{ReLU}(\hat{y}^{t}_{i,0.1} - y^{t}_{i})/\hat{y}^{t}_{i,0.1}
  \label{eq_distance}
\end{equation}
\begin{equation}
  \label{eq_L}
  L^t = \{\frac{1}{I_e-I_s+1}\sum_{i=I_s}^{I_e}\mathcal{L}_q({y^{j}_{i}},\hat{y}^{j}_{i,0.5})\ |\ j\in [t-k,t-1]\}
\end{equation}

Similarly, the prediction errors $L^t\in \mathbb{R}^k$ (Eq.~\eqref{eq_L}) can also reflect the degree of difference between the prediction and truth. Thus, we have the proposals for whether the workload at time $t$ of each service is bursty:
\begin{equation}
  V^t = \{(D^t[i]>\lambda_d\ \text{and}\  O^t[i]\ge n/2)\ \text{or}\ L^t[i]>\lambda_l\ |\ i\in[0,k)\}
  \label{eq_V}
\end{equation}
\replaced[id=REP]{where $\lambda_{l}$ and $\lambda_{d}$ are preset thresholds for controlling the sensitivity of burst detection:}{} \replaced[id=ADD]{
lower thresholds potentially lead to False Positives,  where transient spikes or minor deviations are treated as bursts, resulting in over-provisioning; conversely, higher thresholds may cause False Negatives, where actual bursts are missed, risking SLO violations. We recommend determining these values via a descending search strategy to maximize resource efficiency while prioritizing SLO assurance.}{}

As shown in Algorithm~\ref{alg:burst_detection}, our burst detection algorithm first initializes $D^t_n$,$O^t_n$,$L^t_n$, and $V^t_n$ for each service $n\in[1,N]$ according to the above equations (Lines 1-3).
Then different detection policies are formulated based on the burst state $S^t_n[-1]\in \{1,0\}$ that indicates whether the nearest time step $t-1$ is bursty or not. 
When $S_n^t[-1]==0$, the current workload is considered as burst if two conditions are satisfied: (i) there are proposal(s) suggested as burst in the nearest $n$ proposals $V^t_n[-p:]$; (ii) the nearest outlier information $O^t[-1]$ shows that there are outlier(s) (Lines 4-5).
When $S^t_n[-1]==1$, the confidence of subsequent predictions is reduced by the influence of the burst.
In this case, if either of conditions (i) and (ii) is satisfied, it is considered a burst (Lines 6-7).
If it is still suggested to be non-burst, a further decision on whether it is a burst is made using a majority rule based on the nearest $k$ ($>p$) proposals (Lines 8-11).

\begin{algorithm}[htbp]
  \renewcommand{\algorithmicrequire}{\textbf{Input:}}
	\renewcommand{\algorithmicensure}{\textbf{Output:}}
  \caption{Pseudocode of Burst Detection}
  \label{alg:burst_detection}
  \begin{algorithmic}[1] 
      \STATE Initialize predicted workloads $\hat{\mathbf{Y}}^t$, ground truth workloads $\mathbf{Y}^t$, last $k$ burst state $S^t$, nearest range length $n$,
      thresholds $\lambda_d$ and $\lambda_l$
      \STATE Compute $D^t_n$, $O^t_n$, $L^t_n$ according to Eqs.~\eqref{eq_D},~\eqref{eq_O} and \eqref{eq_L} with input $\hat{\mathbf{Y}}^t$ and $\mathbf{Y}^t$ for each service $n\in[1,N]$
      \STATE Compute
        $V^t_n$ according to Eq.~\eqref{eq_V} with input $D^t_n$, $O^t_n$, $L^t_n$, $\lambda_d$ and $\lambda_l$
      \IF{\NOT $S^t_n[-1]$}
        \STATE Set $B_n=\sum V^t_n[-p:]>0$ and $O^t_n[-1]>0$
      \ELSE
        \STATE Set $B_n=\sum V^t_n[-p:]>0$ or $O^t_n[-1]>0$
        \IF{$B_n$ equals 0}
          \STATE Remove the items in $V^t_n$ with index $i$ satisfying $S^t_n[i]!=1$
          \STATE Set $B_n=\sum V^t_n[-p:] \ge p/2$ 
        \ENDIF
      \ENDIF
    
      \RETURN Result of burst detection $B^t=\{B_n| n\in[1,N]\}$
  \end{algorithmic}
\end{algorithm}

\textbf{Burst Handling.}\label{sec:method_burst_handling}
Since bursts tend to be a tiny part of overall workloads but have a considerable impact, proper overestimation helps to improve the robustness of burst handling. 
Thus the burst handler is designed to provide appropriately overestimated values of bursts. In this component, we first use a second-order AR~\cite{bollerslev1986generalized} to model the estimation of bursts. 
Unlike the Transformer-based workload predictor that provides valuable insights into the data's long-term trends and global patterns, the AR model emphasizes capturing the local features of recent observations, thus better adapting to short-term dynamics during bursts. Specifically, an AR model has dynamics given by:
\begin{equation}
  \hat{y}^{t+1} = c + \phi_1y^{t} + \phi_2y^{t-1} + \epsilon^t
  \label{eq:autoregressive}
\end{equation}
where $y^t=WL^t$ and $y^{t-1}=WL^{t-1}$ are the inputs to the model representing the observed workloads at time $t$ and $t-1$. 
$\hat{y}^{t+1}$ is the model output representing the predicted workload at the future time $t+1$.  
$c$, $\phi_1$, $\phi_2$ and $\epsilon^t$ are the parameters, and $\epsilon_t$ is assumed to be a white noise process. 
For each service $n$, given its most recent $L_z$ observations of workload, the model parameters are optimized using \emph{Maximum Likelihood Estimation}.
Then we add a residual term to the estimation by the AR model to generate an appropriate overestimate for bursts (Eq.~\eqref{eq_y_overestimate}).
\begin{equation}
  \label{eq_y_overestimate}
  \hat{y}^{t+1}_{n,burst} = \hat{y}^{t+1}_n + \operatorname{U}_{\operatorname{95\%CI}}[\operatorname{Q_{95\%}}(E^t_n)]
\end{equation}
where $\operatorname{U}_{\operatorname{95\%CI}}[\cdot]$ signifies the upper bound of the 95\% confidence interval (CI) obtained through the bootstrap procedure, and $\operatorname{Q_{95\%}}(E^t_n)$ represents the 95th percentile of the residuals $E^t_n$ of service $n$. $E^t_n$ is generated by:
\begin{equation}
  E^t_n = \{y^j_n-\hat{y}^j_n\ |\ j\in [t-k+1,t],n\in[1,N]\}
\end{equation}

\replaced[id=ADD]{In effect, integrating the burst detector and handler with the Transformer-based predictor enables a tiered response strategy for diverse burst dynamics. Transient spikes are filtered via sensitivity thresholds (i.e., $\lambda_{l}$ and $\lambda_{d}$) to prevent resource flapping. Short-term bursts are rapidly handled by the AR model via generous resource provisioning, while sustained bursts are naturally absorbed as the Transformer's sliding window updates, ensuring seamless adaptation to new workload baselines.}

\begin{table}[htbp]
  \caption{State, Action and Reward of the DRL agent}
  \begin{center}
    \begin{tabular}{@{}
      >{\columncolor[HTML]{ECECEC}}c 
      >{\columncolor[HTML]{ECECEC}}l @{}}
      \toprule
      \multicolumn{2}{c}{\cellcolor[HTML]{ECECEC}\textbf{State $S_t\in \mathcal{S}$}}                                                                                                             \\
      \multicolumn{2}{c}{\cellcolor[HTML]{FFFFFF}\begin{tabular}[c]{@{}c@{}}Resource Utilization ($RU_t$), Response Time ($RT_t$),\\ Instances Number ($IN_t$),  Workload Embedding ($WE_t$), \\
      Memory Usage ($MU_t$), Service ID Embedding ($SVC$)  
      \end{tabular}} \\
      \multicolumn{2}{c}{\cellcolor[HTML]{ECECEC}\textbf{Action $A_t\in \mathcal{A}^{p_t}_t$}}                                                                                      \\
      \multicolumn{2}{c}{\cellcolor[HTML]{FFFFFF}        Origin Pivot Space $\mathcal{A}^{p_t^0}_t$, Prediction Pivot Space $\mathcal{A}^{p_t^1}_t$
       }                                                                                                      \\
      \multicolumn{2}{c}{\cellcolor[HTML]{ECECEC}\textbf{Reward $R_t\in \mathcal{R}$}}                                                                                                            \\
      \multicolumn{2}{c}{\cellcolor[HTML]{FFFFFF}SLO Assurance Score ($R_t^{RT}$), Resource Efficiency Score ($R_t^{RU}$)}                                                                       \\ \bottomrule
      \end{tabular}
  \end{center}
  \label{tab:state_action_reward}
\end{table}

\subsection{Resource Estimation}\label{sec:method_resource_estimation}
In the resource estimator module, we estimate resource requirements indirectly by combining performance modeling with predicted request workloads. Since both workload prediction and performance modeling rely solely on historical observational data for policy learning, this approach eliminates the need for expensive labeled data in resource estimation.

\textbf{Performance Modeling.} To implement this, we first design a SVR~\cite{awad2015support} driven performance model to estimate the response time under given workloads, instance replicas, and service ID. SVR, a powerful machine learning technique, is well-suited for performance modeling due to its ability to capture non-linear relationships. 
Our performance model is represented as following:
\begin{equation}
  f_{\operatorname{SVR}}(x) = w\phi(x) + b
\end{equation}
where $f_{SVR}(\cdot)$ denotes the predicted response time, $x=\{IN, WL, SVC\}$ contains the workload intensity, number of instances, and service ID embedding (i.e., one-hot encoding of service ID).
$\phi(x)$ denotes the \emph{Radial Basis Function} kernel applied to the input variables $x$, and $w$ and $b$ are the model parameters.

Historical data is collected from the target system to train the SVR model. 
These data consist of various workload scenarios and their corresponding response times.
The workload scenarios cover different levels of workload intensity and resource quantities, with the workloads and resource quantities serving as the input variable and the response time as the output variable. Once trained and validated, the SVR model can predict the response time for new workload scenarios.

\textbf{Performance-Constrained Resource Estimation.} In the context of resource planning, the SVR-based performance model is utilized to determine the minimum resource quantities (i.e., $IN_{min}$) needed to meet an SLO-defined response time requirement $\lambda_{RT}^{n}$. Given the performance model $f_{SVR}$, a predicted workload $WL_{pred}$, and service $n$ (i.e., SVC), the performance-constrained resource estimation can be expressed as the following: 
\begin{equation}
  \begin{aligned}
    IN_{min} = \min_{IN}\ \ s.t.\ f_{\operatorname{SVR}}(\{IN, WL_{pred},SVC\})<\lambda_{RT}^{n}
  \end{aligned}
  \label{eq:resource_estimate}
\end{equation}
where $WL_{pred}=\hat{y}^{t+1}_{n,burst}$ for bursts and $WL_{pred}=\hat{y}^{t}_{1,n,0.5}$ for non-bursts.
By iteratively adjusting the resource quantities and observing the corresponding predicted response times, we identified the optimal resource configuration that satisfies the response time constraint.

\subsection{Estimation Enhancement}
\label{sec:method_estimation_enhance}
Compared to burst situations that require an appropriate overestimation strategy, non-burst situations require a more refined resource estimate. Thus, we design a DRL agent, based on \emph{Proximal Policy Optimization} (PPO)~\cite{schulman2017proximal}, to correct possible misestimations by the above method for non-bursts due to its ability to learn from experience and adapt dynamically to changing conditions. 
Specifically, at each discrete time step $t=1,2...$, our DRL formulation modeled as a \emph{Markov Decision Process} $(\mathcal{S},\mathcal{A},T,\mathcal{R},\gamma)$, where $\mathcal{S}$ is a state space, $\mathcal{A}$ is a set of actions available to an agent, $T(s,a,s^{'})$ is the (stochastic) transition function, $\mathcal{R}$ is a reward space defined by reward function and $\gamma\in[0,1]$ is the discount factor. 
The goal of the DRL agent is to optimize policy $\pi_\theta$ to maximize the expected cumulative reward. 

Table~\ref{tab:state_action_reward} shows the state, action, and reward design of BASE at time $t$. The state $S^t \in \mathcal{S}$ is defined as a tuple $(RU^t,RT^t,IN^t,WE^t,MU^t,SVC)$.

\replaced[id=REP]{It contains the average resource utilization of CPU $RU^t$ and response time $RT^t$ aggregated across all replicas, the total number of instances $IN^t$, the workload embedding $WE^t$, the average memory usage $MU_t$, and Service ID embedding $SVC$ at the timestep $t$.
$WE^t$ contains the workload representation (described in §\ref{sec:method_wkl_predict}) for the last $k$ timesteps.
The action $A^t \in \mathcal{A}$ constitutes a service-level scaling decision, determining the target total number of instances for the service. It is fine-tuned based on the predicted instance number $p_0^t=IN_{min}$ and current instance number $p_1^t=IN^{t}$:}{} 
\begin{equation}
  \mathcal{A}=\mathcal{A}_{p^t_0}^t \cup \mathcal{A}_{p^t_1}^t;\ \mathcal{A}_{p^t_i}^t=\{p^t_i,p^t_i\pm 1,p^t_i\pm 2,...,p^t_i\pm \sigma\};\ i\in \{0,1\}
  \label{eq:action}
\end{equation}
where $\mathcal{A}^t_{p_i^t}$ represents the sub-action space based on $p_i^t$, 
$\sigma$ is used to control the fine-tuning range of the instance number.
The reward $R^t = \beta R^t_{RU} + (1-\beta) R^t_{RT}$ is defined as a linear combination of $R^t_{RT}$ and $R^t_{RU}$ by using hyperparameter $\beta \in (0,1)$, where $R^t_{RT}$ and $R^t_{RT}$ are designed to measure SLO assurance and resource efficiency:
\begin{equation}
  R^t_{RU}=\left\{
  \begin{aligned}
   & RU^t/\lambda_{RU},\ \ \ \operatorname{if} RU^t\leq \lambda_{RU} \\
   & 2-\frac{1-\lambda_{RU}}{1-RU^t},\ \ \ \ \ \ \operatorname{else}
  \end{aligned}
  \right.
  \label{eq_reward}
\end{equation}
\begin{equation}
  R^t_{RT}=\left\{
    \begin{aligned}
     & 1,\ \ \ \ \ \ \ \ \ \ \operatorname{if} RT^t \leq \lambda_{RT} \\
     & \alpha(\frac{\lambda_{RT}}{RT^t}-2),\ \ \ \ \  \operatorname{else} 
    \end{aligned}
    \right.
\end{equation}
where $\lambda_{RU}$ represents a CPU utilization threshold above which the QoS deteriorates rapidly, and $\alpha$ is a penalty factor for response time violations.

\replaced[id=ADD]{\textbf{Why no DRL for Bursts?} It is worth to say that restricting DRL to non-burst scenarios reflects a trade-off between Safety and Optimization. Since DRL agents inherently require an exploration process and stable state distributions to converge. However, in critical, uncertain, and infrequent burst scenarios, such exploration not only impedes convergence but also incurs high risks of under-provisioning. Conversely, the simpler AR mechanism combined with Bootstrapping functions as a deterministic "safety net." 
By employing generous resource provisioning, it prioritizes immediate service availability.}{}

\subsection{\replaced[id=ADD]{Deployment and Evolution}}\label{sec:method_policy_learning}

\replaced[id=ADD]{
To ensure practical manageability, BASE follows a three-stage lifecycle pipeline. The process begins with Service Onboarding, designed to address the cold-start problem where historical data is absent. In this stage, BASE operates in a fallback mode, utilizing industry-standard scaling (i.e., Kubernetes HPA) to ensure basic stability while aggregating high-resolution telemetry data. 
Once sufficient historical data is available, it is used to initialize the Transformer, AR, and SVR models. 
}{}

\replaced[id=ADD]{
Upon initialization, the system transitions to Online Execution. In this active phase, the lightweight components adapt in real-time: the DRL agent continuously updates its policy parameters via online PPO learning during non-burst periods, and the AR model dynamically refits using a sliding window of recent observations. This design allows BASE to adjust to immediate workload changes and refine its decision-making logic without requiring constant manual intervention or heavy offline computation for every update.
}{}

\replaced[id=ADD]{
Finally, to address long-term concept drift and environment variability, BASE incorporates Periodic Maintenance. The Transformer model is retrained asynchronously on a periodic basis (e.g., weekly) to capture evolving workload patterns, while the SVR performance model is similarly updated to adapt to potential shifts in the resource-performance relationship (e.g., due to application updates or infrastructure changes). These updated models are deployed only after achieving a lower validation loss than the current active versions. This tiered approach ensures that the stacked ensemble maintains high accuracy and robustness throughout the application lifecycle.
}{}

\replaced[id=ADD]{Notably, although fine-grained hyperparameter tuning can enhance performance, BASE adopts a hybrid strategy to balance tuning complexity and adaptability. Structural hyperparameters (e.g., Transformer architecture) are initialized with fixed values, leveraging their empirical robustness. In contrast, burst detection thresholds are calibrated via the descending search strategy to adapt to different environments.}{}

\section{Experimental Evaluation}
In this section, we evaluate {BASE} to answer the following questions:
\begin{itemize}

    \item \textbf{RQ1:} How effective is BASE? 
    \item \textbf{RQ2:} How adaptive is {BASE}? 
    \item \textbf{RQ3:} What is the contribution of each design in BASE?
\end{itemize}

\subsection{Experimental Settings}\label{sec:eval_settings}
{\textbf{System Setup.}} 
The experiments are conducted on a Kubernetes cluster deployed on a public \emph{Elastic Compute Service} (ECS) platform, running Kubernetes v1.23.4. The testbed cluster comprises eight VMs running Ubuntu 18.04 LTS. Four VMs have 24 CPU cores and 32~GB of memory, while the remaining four have 12 CPU cores and 24~GB of memory. Istio service mesh v1.14.3 is employed to manage network traffic and provide load balancing.

\textbf{Benchmark Applications.}
We deployed five benchmark applications to evaluate BASE. (i) DFT~\cite{DFT}: A lightweight function service that computes and outputs the discrete Fourier transform and its inverse for digital signals. (ii) Bookinfo~\cite{bookinfo}: provided by Istio, is an online bookstore application that includes detailed information about a book. (iii) Online-Boutique~\cite{boutique}: a cloud-native microservices application used by Google to showcase the functionality of Kubernetes/GKE, Istio, and gRPC.
This web-based e-commerce application allows users to browse a range of merchandise, add items to their cart, and complete their purchases. 
(iv) Sock-Shop~\cite{sockshop}: the user-facing part of an online shop that sells socks build using Spring Boot, Go kit and Node.js. (v) Train-Ticket~\cite{trainticket}: a ticket booking application comprising 41 microservices, each responsible for a specific function, such as user authentication, ticket booking, payment processing, and notification, for a comprehensive evaluation in a multifunctional scenario. 

\textbf{Workload Datasets.} We employed \emph{Grafana K6}~\cite{grafana_k6}, an open-source load testing tool, to generate concurrent digital signals, simulating specific workload traces to drive the service. Different workload traces were used to represent diverse active production environments, including ten different real-world workload traces publicly available at \cite{wikimedia_rest_api} and maintained by the \emph{Wikimedia Foundation}, including the pageviews of \emph{Barack Obama}, \emph{Game of Thrones}, \emph{Donald Trump}, \emph{Elon Musk}, \emph{2018 FIFA World Cup}, \emph{Facebook}, \emph{Elizabeth II}, \emph{United States}, \emph{YouTube}, and \emph{Google}. These workloads encompass almost all the characteristics of real-world usage patterns, such as burstiness, randomness, trend, and seasonality.

\begin{table}[htbp]
  \centering
  \caption{Implementation Details}
  \begin{tabular}{c|c|l}
  \toprule[1pt]
  Component& Impl. & Parameter Settings                 \\ \midrule

  \begin{tabular}{@{}c@{}}Workload Predictor  \\ (§\ref{sec:method_wkl_predict})\end{tabular} & Informer~\cite{informer_impl}  & $L_x$=720, $L_y$=168 \\ \midrule
  \multirow{3}{*}{\begin{tabular}{@{}c@{}}Burst Detector \& \\ Burst Handler \\ (§\ref{sec:method_burst_detection_and_handling})\end{tabular}} & Algo.~\ref{alg:burst_detection} & $k$=24,$n$=3,$\lambda_d$=0.1,$\lambda_l$=0.1 \\

& AR~\cite{ar_impl} & $\phi_1$=2, $\phi_2$=-1, $L_z$=168 \\

& Bootstrapping~\cite{boostrapping_impl} & iteration number = 100 \\ \midrule

\begin{tabular}{@{}c@{}}Resource Estimator  \\ (§\ref{sec:method_resource_estimation})\end{tabular} & SVR~\cite{svr_impl} & kernel="rbf", $C$=100 \\ \midrule

\multirow{2}{*}{\begin{tabular}{@{}c@{}}Estimation \\ Enhancement (§\ref{sec:method_estimation_enhance}) \end{tabular}} & \multirow{2}{*}{PPO~\cite{ElegantRL}} & $\sigma$=2, $\beta$=0.5, $\lambda_{RU}$=0.9  \\
& & $\gamma$=0.99, $\lambda_{RT}$=16ms, $\alpha$=1 \\ \bottomrule[1pt]
  \end{tabular}
  \label{tab_parameter}
\end{table}

\textbf{Implementation Details.}
We implemented BASE using Python 3.8, incorporating the implementation details and parameter settings outlined in Table~\ref{tab_parameter}. Each component's underlying model is built upon open-source code, with references listed in the table. We trained the improved \emph{Transformer} model, SVR model, and RL agent using their respective components' APIs. Specifically, the \emph{Informer} model was trained on the first 10,000 hours of data from ten benchmark workload traces, while the SVR model was trained using historical testbed data. The RL agent was trained through interactive interactions with the testbed.

\textbf{Baseline Autoscalers.} We implemented the following baselines for comparison: (1) \textbf{\emph{HPA}}~\cite{k8sHPA}, Kubernetes' default horizontal pod autoscaler, which scales resources based on CPU utilization and predefined thresholds, representing the most widely used rule-based reactive approach; (2) \textbf{\emph{A-SARSA}}~\cite{zhang2020sarsa}, which uses an RL agent with a fused prediction network to determine scaling actions; we select \emph{A-SARSA} because it exemplifies an ensemble approach that integrates multiple ML models; (3) \textbf{\emph{BAPA}}~\cite{abdullah2020burst}, a burst-aware autoscaling approach that detects bursts in real time using statistical methods and employs decision trees for resource estimation, representing a hybrid method incorporating burst detection; (4) \textbf{\emph{DeepScaler}}~\cite{DeepScaler}, a deep learning--based autoscaling method that models microservice dependencies via attention-based graph convolution, chosen as a representative of dependency-adaptive proactive autoscaling methods; and (5) \textbf{\emph{ARAScaler}}~\cite{ARAScaler}, an adaptive autoscaling method that segments predicted workloads using ETimeMixer and adjusts resources by workload state, selected as a representative of state-aware autoscaling methods.

\begin{figure}[htbp]
	\centering
	\includegraphics[width=\linewidth]{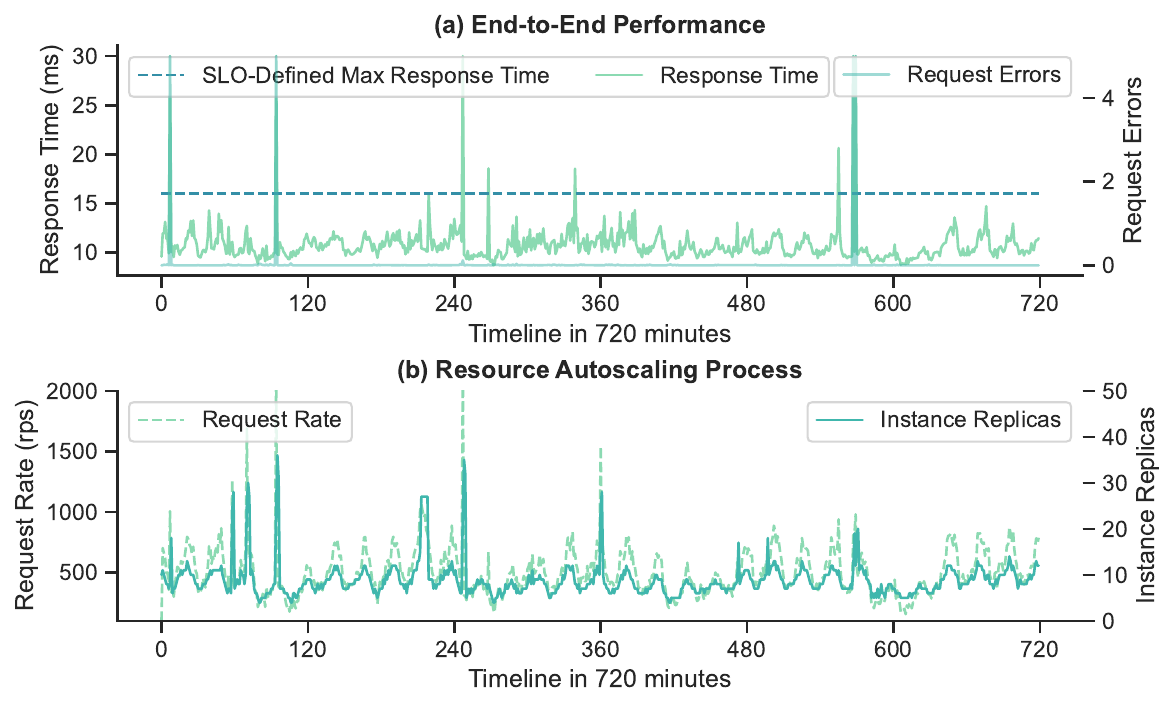}
	\caption{Autoscaling Process of BASE on Google Workload}
	\label{fig_eval_autoscaling_efficient}
\end{figure}

\textbf{Evaluation Metrics.} 
We evaluate all methods using the following metrics: SLO violation, resource cost, request error, resource scaling score, response time variance, SLO violation variance, and resource cost variance.
The SLO violation measures the number of time steps during which the SLO is not met. Resource cost reflects the total number of instances consumed, and request error count indicates the number of failed requests during execution.
The scaling score, defined in Eq.~\ref{eq_score}, quantifies the score of resource scaling decisions. The three variance-based metrics are used to capture the stability of the system.
For all metrics, lower values indicate better performance, except for the scaling score, where a higher value is preferred.

\begin{equation}
  Score = 1 - \frac{1}{N} \sum_{n=1}^{N} \left( \frac{|RT_n - \lambda_{RT}^n|}{\lambda_{RT}^n} \right)
  \label{eq_score}
\end{equation}

\subsection{Overall Evaluation (RQ1)}\label{sec:eval_effectiveness}
\textbf{Effectiveness.} In accordance with recent workloads and telemetry indicators, BASE dynamically reassesses the resource demands and allocates them accordingly at one-minute intervals. Fig.~\ref{fig_eval_autoscaling_efficient} illustrates a 12-hour run (720 time steps) of BASE on the Google trace. In Fig.~\ref{fig_eval_autoscaling_efficient}(a), the observed response time almost always stays below the SLO threshold; only a few $40\times$ demand bursts briefly push it higher, yet the tail latency quickly converges without prolonged errors. Correspondingly, Fig.~\ref{fig_eval_autoscaling_efficient}(b) shows that BASE copes with incoming request changes through adaptive replica adjustments. 
These facts confirm that BASE's burst-adaptive resource planning coupled with RL calibration implements an effective autoscaling mechanism that can maintain a good balance between QoS and resource cost in dynamic workloads.

\begin{figure}[htbp]
	\centering
	\includegraphics[width=0.85\linewidth]{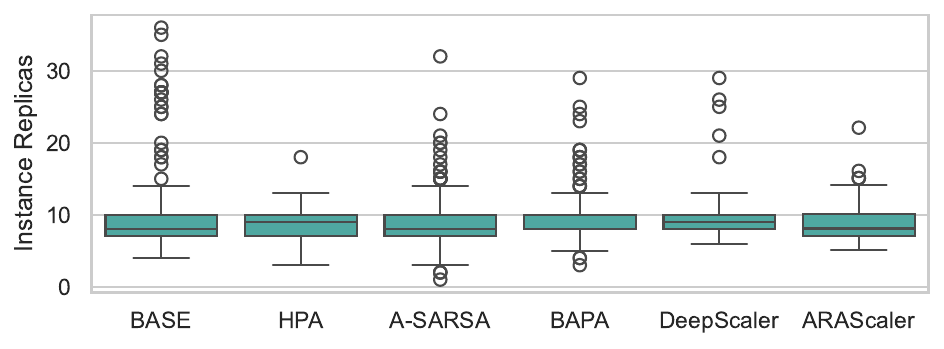}
	\caption{Distribution of Instance Consumption}
	\label{fig_eval_autoscaling_google_cost}
\end{figure}
\textbf{Comparisons.} To make an apple-to-apple comparison between BASE and other baseline methods, we carefully controlled the resources allocated to the application by the different methods at similar levels. 
As shown in Fig.~\ref{fig_eval_autoscaling_google_cost}, all six methods operate under nearly identical resource envelopes—their box-plots share similar medians ($\approx$ 8-9 replicas) and inter-quartile rangeThe comparative results are presented in Table~\ref{tab_real}, where BASE demonstrates superior performance across all metrics, achieving significant improvements ranging from 59\% to 71\%. 
We argue that BASE's outperformance over other methods is due to the following factors: (i) its proactive scaling based on predictions is more timely than reactive scaling; (ii) the ensemble strategy, which integrates multiple models (including RL) for resource estimation, yields greater accuracy than single-model predictions; and (iii) the burst detection and handling mechanism effectively addresses workload bursts, ensuring robust performance under burst scenarios.

\begin{figure*}[htbp]
  \centering
  \includegraphics[width=\linewidth]{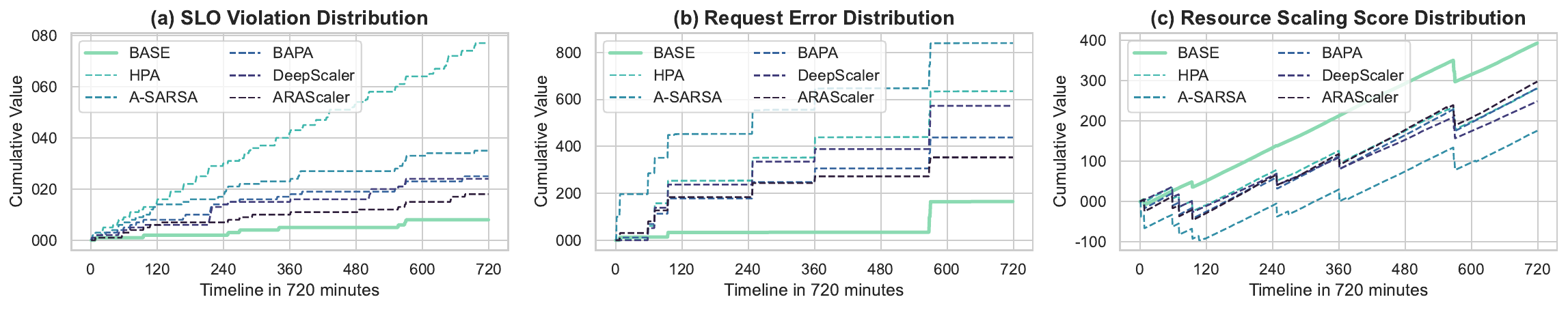}
  \caption{Performance Comparison over Time on \emph{Google} Workload
  }
  \label{fig:eval_real_distribution}
\end{figure*}

\begin{table}[htbp]
  \centering
  \caption{Quantitative result of the effectiveness evaluation}
  \label{tab_real}
  \begin{tabular}{@{}c|c|c|c|c@{}}
    \toprule[1pt]
    \multirow{2}{*}{ Autoscalers} & \multicolumn{4}{c}{Metrics}   \\ 
    \cmidrule(lr){2-5} 
    & Violation ($\downarrow$)  & Error ($\downarrow$) & Score ($\uparrow$) & Var. ($\downarrow$) \\ 
    \midrule
    \textbf{BASE} & \textbf{8.00} & \textbf{164.91} & \textbf{392.75} &\textbf{718.44}\\ 
    HPA & 77.00 & 635.14 & 280.65 &2134.27\\
    A-SARSA & 35.00 & 840.04 & 175.53 & 3193.30\\
    BAPA  & 25.00 & 437.96 & 281.16 &2426.69 \\
    DeepScaler & 24.00 & 572.99 & 248.49 &2067.74\\ 
    ARAScaler & 18.00 & 353.26 & 297.45 &2246.04\\ 
    \midrule
    \textbf{Improvement} & \textbf{71\%$\pm$11\%} & \textbf{68\%$\pm$9\%} & \textbf{59\%$\pm$34\%} & \textbf{70\%$\pm$4\%}\\
    \bottomrule[1pt]
    \end{tabular}
\end{table}

Moreover, Fig.~\ref{fig:eval_real_distribution} shows the cumulative distribution of SLO violation, request error, and resource scaling score over time. 
As evidenced by the figure, the performance benefits of BASE exhibit a cumulative effect over time, suggesting that the longer BASE is equipped, the greater the benefits it yields. This is particularly valuable in continuous operation environments, where BASE ensures better service quality and significantly reduces resource manage costs.

\subsection{Adaptability Evaluation (RQ2)}
\label{sec:eval_adaptability}

\textbf{Varying Different Workloads.} Here, we evaluate the adaptability of the proposed autoscaler under larger-scale settings—with extended time series up to 4000 time steps—and more diverse workloads.
To evaluate workload-level adaptability, we replayed ten heterogeneous production-like traces—ranging from burst-heavy workloads (FIFA World Cup, YouTube) to more quiescent patterns (Elizabeth II, Donald Trump)—each extended to 4000 time-steps. 

\begin{figure}[htbp]
  \centering
  \includegraphics[width=\linewidth]{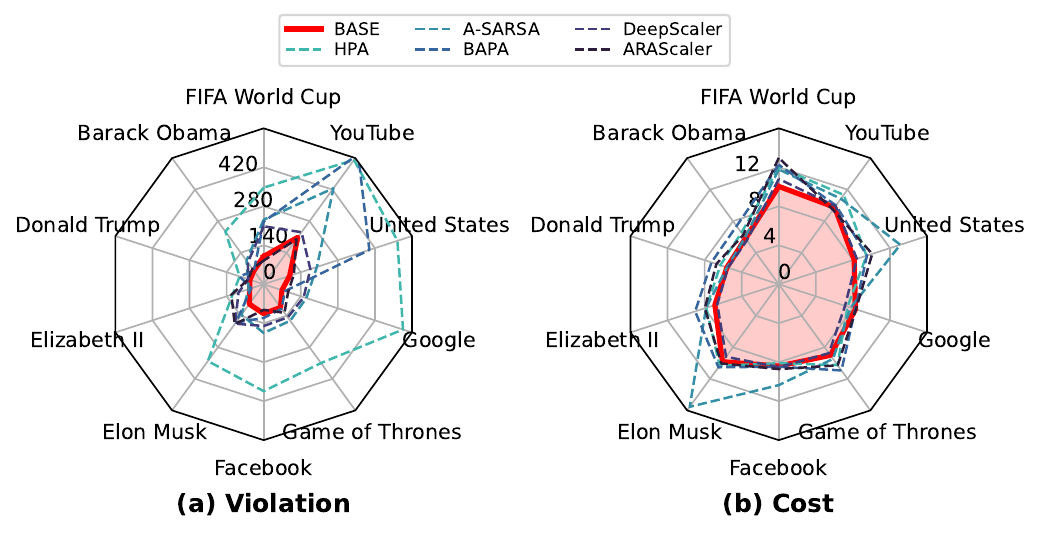}
  \caption{Aaptability Evaluation on Ten Workloads
  }
  \label{fig:eval_sim_radar}
\end{figure}

Fig.~\ref{fig:eval_sim_radar} provides a detailed comparison across the ten workloads.
In the SLO-violation radar (Fig.~\ref{fig:eval_sim_radar}a) BASE consistently forms the smallest polygon, demonstrating uniformly low violation counts across every workload, whereas HPA and A-SARSA either over- or under-provision in highly variable traces and DeepScaler/ARAScaler still miss deadlines when their dependency or state models are less reliable. 
The replica-count radar (Fig.~\ref{fig:eval_sim_radar}b) shows that BASE achieves these QoS gains without inflating resource usage; its allocation profile remains compact and balanced, unlike the aggressive over-provisioning of HPA or the high variance of A-SARSA.
The dispersion statistics in Table~\ref{tab_simulation} quantify this stability: BASE reduces SLO-violation variance to 1853.41, 
an average reduction of 85.7\% compared to the competing methods, and lowers cost (replica) variance to 2.31, a 45.0\% improvement. These twin reductions indicate that BASE maintains predictable QoS while avoiding oscillatory or overly conservative scaling, validating its capacity to generalise to different workload shapes with minimal performance or cost degradation.

\begin{table}[htbp]
  \centering
  \caption{Quantitative result of the adaptability evaluation}
  \label{tab_simulation_transposed}
  \resizebox{\linewidth}{!}{ 
    \begin{tabular}{
      @{}
      >{\centering\arraybackslash}p{1.55cm}  
      |>{\centering\arraybackslash}p{1cm}
      |>{\centering\arraybackslash}p{1cm}
      |>{\centering\arraybackslash}p{1.3cm}
      |>{\centering\arraybackslash}p{1.2cm}
      |>{\centering\arraybackslash}p{1.4cm}
      |>{\centering\arraybackslash}p{1.4cm}
      @{}
    }
      \toprule[1pt]
      \multirow{3}{*}{Metrics} & \multicolumn{6}{c}{Autoscalers} \\ 
      \cmidrule(lr){2-7} 
                     & \textbf{BASE}        & HPA                         & A-SARSA               &    BAPA                   & DeepScaler              &  ARAScaler               \\ 
      \midrule
      Vio. Var. ($\downarrow$) & \textbf{1853.41} & \emph{23159.61} & \emph{10924.24} & \emph{26176.05} & \emph{2551.96} & \emph{1987.05} \\       \\
      Cost Var. ($\downarrow$) & \textbf{2.31} & \emph{3.61} & \emph{9.03} & \emph{2.33} & \emph{2.87} & \emph{3.42} \\
      \midrule
      Improvement & \multicolumn{6}{c}{\textbf{85.70\%} on SLO violation variance, \textbf{45.00\%} on cost variance} \\
      \bottomrule[1pt]
    \end{tabular}
  }
  \label{tab_simulation}
\end{table}

\begin{figure}[htbp]
  \centering
  \includegraphics[width=0.9\linewidth]{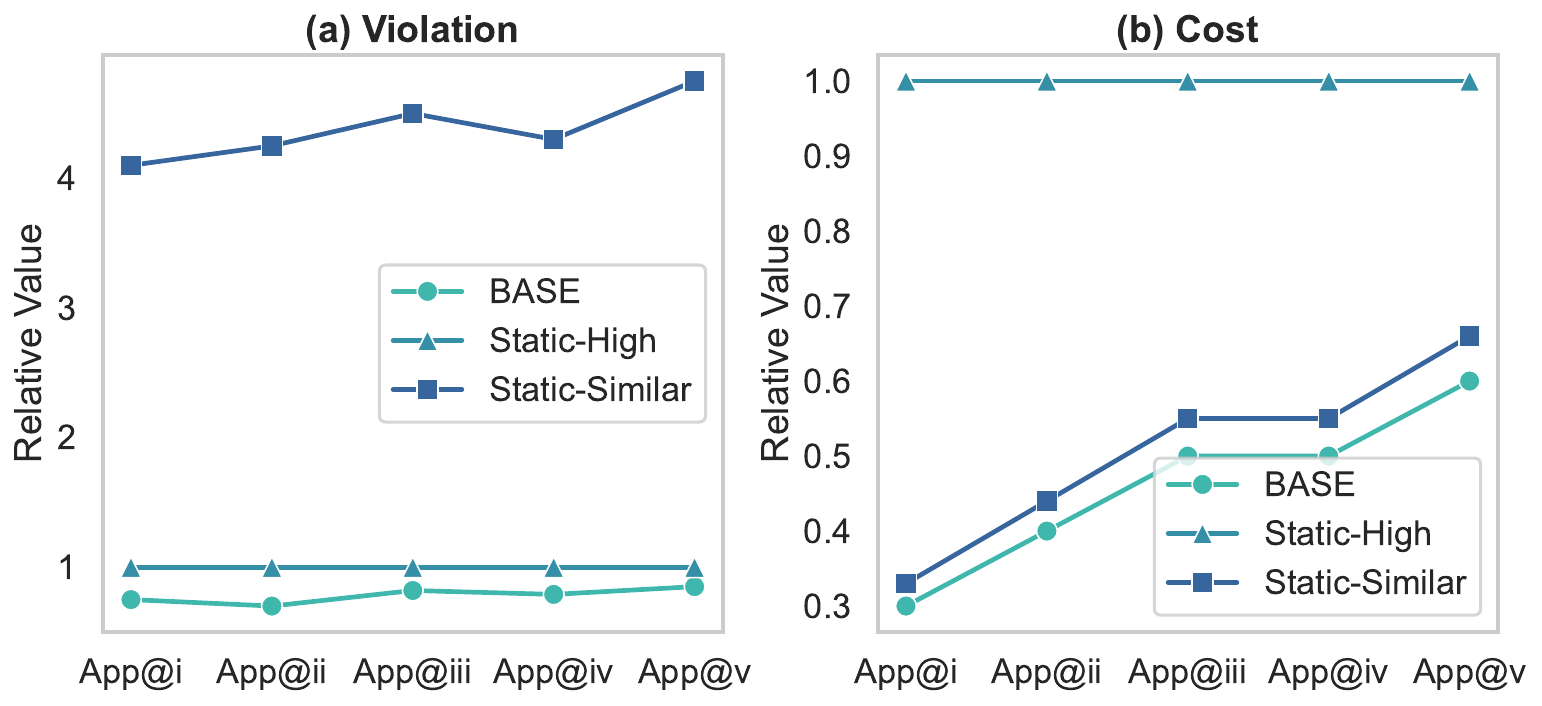}
  \caption{Adaptability Evaluation on Five Microservice Applications. 
  }
  \label{fig:eval_diff_app}
\end{figure}

\textbf{Varying Different Applications.} To evaluate the generalizability of our proposed approach across a variety of microservice applications, we conducted experiments on five widely adopted benchmarks: (i) DFT, (ii) Bookinfo, (iii) Online-Boutique, (iv) Sock-Shop, and (v) Train-Ticket. We compared our adaptive autoscaling method against two static resource allocation strategies. The first strategy, denoted Static-Similar, allocates resources equal to 110\% of those used by the baseline. The second strategy, Static-High, provisions resources sufficient to satisfy the 95th percentile peak load.
Fig.~\ref{fig:eval_diff_app} summarises the cross-benchmark study. For every application (App@i-App@v), BASE delivers the lowest SLO Violation and the lowest replica cost. In contrast, Static-Similar incurs 5-6x more violations, while Static-High consumes 2-3x the resources to maintain a comparable QoS level. These results underscores that BASE generalises across heterogeneous microservice stacks, consistently striking a superior balance between SLO guarantees and resource efficiency compared with static provisioning schemes.

\textbf{Analysis.} BASE's consistent superiority across ten workload datasets and five benchmark applications validates its adaptability and generalizability. 
We argue that these benefits stem from BASE's two core designs: (i) \emph{Application-Architecture-Agnostic:} The data-driven Transformer implicitly captures workload propagation without static graphs, while independent SVR models adapt to heterogeneous service logic and resource patterns; and (ii) \emph{Workload-Pattern-Agnostic:} 
The integration of workload forecasting and burst handling manages both expected fluctuations and unexpected bursts, while real-time DRL updates and periodic retraining jointly address short-term deviations and long-term concept drift.

\subsection{Ablation Study (RQ3)}\label{sec:eval_ablation_study}

To analyze the contribution of each design in BASE, we conducted ablation studies building upon the foundation of adaptability evaluation experiments (§\ref{sec:eval_adaptability}). By selectively omitting specific modules of the proposed approach, we introduce three new autoscaling methods, which are as follows:
\begin{itemize}
  \item \textbf{BASE w/o Burst}: An ablation method for the burst detection and handling module, where all workloads allocate resources solely based on resource estimation and the estimation enhancement module.
  \item \textbf{BASE w/o Pred}: Ablation of the resource prediction module, implying an improvement in the reinforcement learning action space by not using resource prediction results in the RL action space, solely relying on the action space based on the current instance axis.
  \item \textbf{BASE w/o RL}: An ablation method for the reinforcement learning-based estimation enhancement module, allocating resources only based on the prediction results from the resource estimation module.
\end{itemize}

\begin{figure}
  \centering
  \includegraphics[width=\linewidth]{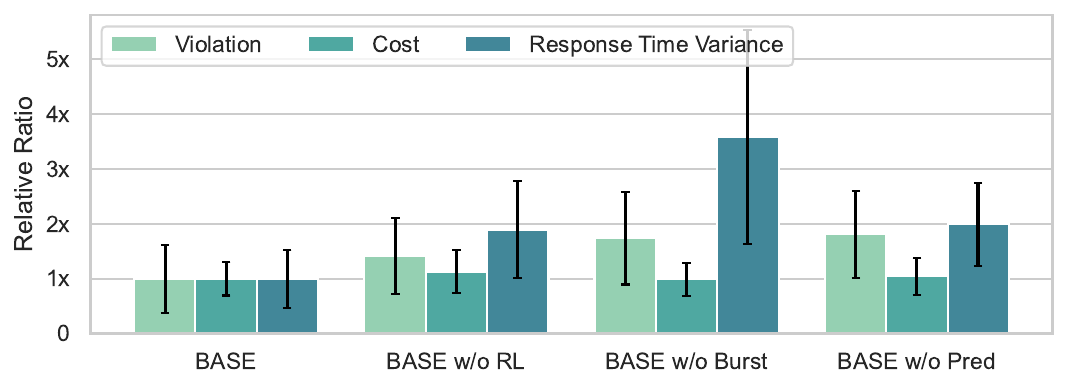}
  \caption{Performance comparison of ablation methods.}
  \label{fig_eval_ab}
\end{figure}

Fig.~\ref{fig_eval_ab} presents the results of the ablation experiments.
From the figure, we observed that BASE outperforms all ablation methods, indicating the effectiveness of each design in BASE. 
Specifically, (i) BASE w/o Burst exhibited a 1.7x increase in SLO violation and attained the highest response time variance at 3.6x, implying that the burst detection and handling module effectively manages bursty workloads, thereby controlling excessive response time and severe fluctuations. Additionally, BASE w/o Burst shows a slight 1\% reduction in resource cost, attributed to the ablation of the burst handling module which adopting a more relaxed resource allocation strategy for bursty workloads.
(ii) BASE w/o Pred exhibited a 1.8x increase in SLO violation, a 2.0x increase in response time variance, and a marginal 4\% increase in resource cost. It means that BASE's resource prediction module effectively predicts resources, guiding more efficient resource allocation.
(iii) BASE w/o RL demonstrated a 1.4x increase in SLO violation, a 1.9x increase in response time variance, and a 1.3x increase in resource cost. It implies the presence of some inaccuracies in resource estimation in BASE, and the reinforcement learning-based estimation enhancement module effectively corrects these erroneous estimates. It is attributed to the online learning capability of reinforcement learning, enabling real-time model adjustment based on the system's state. Compared to offline-trained resource estimation models, reinforcement learning proves more adept at adapting to dynamic changes in the system's environment.

\section{Discussion}
\textbf{Overhead.} 
Although BASE integrates multiple ML models, its deployment overhead does not increase because they execute sequentially; \replaced[id=REP]{consequently, the container hosting BASE is only required to satisfy the peak resource demand of the largest single model (i.e., the Transformer), rather than the aggregate capacity of the ensemble.
Moreover, frequent scaling actions can induce system oscillations, so the scaling interval is fixed at one minute. 
To quantify the computational latency within this interval, we profiled the complete decision process on a container restricted to 0.1 vCPU. Even accounting for the overhead of sequential model loading and inference, the total execution time averages less than 500 ms. This constitutes less than 1\% of the decision cycle, confirming that the stacked ensemble imposes no bottleneck on real-time responsiveness. Consequently, BASE operates reliably with minimal resource allocation. As shown in Table~\ref{tab:overhead}, incurring an estimated monthly overhead of approximately USD 2.8 on AWS Fargate (US-East), BASE provides substantial service-level improvements with effectively negligible operational costs.}{}

\begin{table}[ht]
  \centering
  \caption{Resource Requests and Estimated Monthly Overhead}
  \label{tab:overhead}
  \makebox[\linewidth][c]{%
  \begin{tabular}{c|c|c|c}
    \toprule
    {Resource}       & {Request}  & {Unit Price}        & {Overhead} \\
    \midrule
    CPU (vCPU)       & 0.10       & 0.03238 \$/vCPU·h   & $\approx$2.84\$            \\ 
    Memory (GB)     & 0.20       & 0.00356 \$/GB·h    & per month          \\
    \bottomrule
  \end{tabular}}
  \vspace{1ex}%
  \begin{tablenotes}[flushleft]
    \footnotesize
    \item[*]\emph{Note: Assumes AWS Fargate (US-East) pricing, 30 days/month × 24 h/day, continuous operation; monthly overhead = request × hours × unit price.}
  \end{tablenotes}
\end{table}

\textbf{Limitations and Future Work.} 
While BASE eliminates the need for expert-annotated training data by relying solely on historical telemetry, the effectiveness of its predictive models in capturing long-term workload patterns or resource-performance relationships still depends on the availability of sufficient historical data. This dependence may limit its performance in data-scarce environments.
Moreover, although BASE demonstrates strong adaptability to external workload variations and bursty traffic through dynamic resource provisioning, it lacks mechanisms to detect or mitigate QoS degradation caused by internal system anomalies, such as network latency fluctuations or out of memory errors. Enhancing BASE with online anomaly detection and mitigation mechanisms is a promising direction for future research.
Lastly, this study primarily focuses on autoscaling cloud-native applications deployed in centralized cloud infrastructures. Extending its applicability to other deployment paradigms—such as highly heterogeneous, decentralized, or multi-cloud settings—remains an open challenge.
In future work, we will explore integrating geographically distributed and heterogeneous resources to improve BASE's adaptability in edge and hybrid cloud environments, enabling more robust and cost-efficient autoscaling.

\section{Related Work}
Autoscaling is an active research area in various cloud systems (e.g., databases~\cite{9671185}, microservices~\cite{9740415,zargarazad2023auto}, and web applications~\cite{9328525}). Recent surveys~\cite{qu2018auto,gari2021reinforcement,zhong2022machine,verma2021auto,al2017elasticity} have successfully applied autoscaling methods based on heuristics, queueing theory, control theory, and machine learning to their respective systems.
Depending on the scaling timing, current autoscaling approaches can be categorized into reactive and proactive methods.

\textbf{Reactive methods} focus on scaling resources in response to changes in workload or system conditions. These methods typically use threshold-based policies to trigger scaling actions when certain conditions are met.
Major cloud platforms, such as \emph{Amazon Web Services} and \emph{Google Cloud}, use this technique.
One example is Microscaler proposed by Yu et al.~\cite{yu2020microscaler}, which collects QoS metrics in the microservice infrastructure and utilizes a novel criterion called service power to determine the scaling-needed services. 
By combining online learning and a step-by-step heuristic approach, Microscaler achieves precise and optimal scaling decisions.
While reactive methods excel in gradual and smooth workload changes, they may struggle with drastic fluctuations, resulting in under-provisioning and degradation of QoS.

\textbf{Proactive methods} enable proactive provisioning or de-provisioning of resources by anticipating future application needs~\cite{9763776}. These methods leverage various techniques, such as time series analysis, deep learning, and reinforcement learning, to predict workload or resource usage and make scaling decisions in advance. 
Wang et al.~\cite{10.1145/3542929.3563469} introduced DeepScaling, a proactive autoscaling approach that uses a three-step design. It combines reinforcement learning for resource prediction with two deep neural networks for workload forecasting and CPU utilization estimation. By incorporating these components, DeepScaling ensures stable resource utilization around a refined target level. However, the effectiveness of proactive methods heavily relies on accurate predictions and is limited to predictable workloads with recurring patterns, often struggling to handle random bursts commonly observed in certain applications.

\textbf{Hybrid Methods.} 
\replaced[id=REP]{Some researchers have focused on the limitations of purely reactive or proactive methods, proposing hybrid approaches that combine the strengths of both.
For example, 
Yan et al.~\cite{yan2021hansel} proposed HANSEL, which employs a reinforcement learning controller to dynamically switch between a Bi-LSTM-based proactive module and a Kubernetes HPA-based reactive moduleCheng et al.~\cite{cheng2023proscale} introduced a hybrid offloading strategy that reactively handles overloads while proactively provisioning edge microservices based on prediction results.}{} 
\replaced[id=ADD]{To further tackle the complexities of microservices, Qiu et al. proposed FIRM~\cite{FIRM} to address dependency bottlenecks via critical path analysis and subsequently advanced the field with AWARE~\cite{AWARE} by bootstrapping RL agents with offline policies.}{}
\replaced[id=REP]{Compared to these approaches, BASE uniquely integrates a mechanism to proactively detect and handle workload bursts, while leveraging DRL to refine resource predictions, enabling more accurate allocation and enhanced robustness.}{}

\section{Conclusion}
In this paper, we present BASE, a burst-adaptive autoscaling framework, to manage the resource allocation of containerized cloud services and applications on dynamic workloads for SLO assurance and cost efficiency. With the novel burst detection mechanism and multi-level ML models, BASE can better resolve bursts and estimate resources at a finer granularity. Compared with the SOTA autoscaling approaches, BASE obtains a more efficient resource allocation policy and significantly reduces SLO violations at a lower cost. 


\ifCLASSOPTIONcompsoc
  \section*{Acknowledgments}
\else
  \section*{Acknowledgment}
\fi

\replaced[id=REP]{This work was supported by the NSFC-Guangdong Joint Fund Project (U20A6003), the National Natural Science Foundation of China (61972427), the Research Foundation of Science and Technology Plan Project in Guangdong Province (2020A0505100030), the Natural Science Foundation Innovation and Development Joint Fund of Chongqing (CSTB2022NSCQ-LZX0074 and CSTB2025NSCQ-LZX0077), and the China Postdoctoral Science Foundation (2025MD784116).}{}

\ifCLASSOPTIONcaptionsoff
  \newpage
\fi

\bibliographystyle{splncs04}
\bibliography{ref}

@inproceedings {AWARE,
author = {Haoran Qiu and Weichao Mao and Chen Wang and Hubertus Franke and Alaa Youssef and Zbigniew T. Kalbarczyk and Tamer Ba{\c s}ar and Ravishankar K. Iyer},
title = {{AWARE}: Automate Workload Autoscaling with Reinforcement Learning in Production Cloud Systems},
booktitle = {2023 USENIX Annual Technical Conference (USENIX ATC 23)},
year = {2023},
isbn = {978-1-939133-35-9},
address = {Boston, MA},
pages = {387--402},
url = {https://www.usenix.org/conference/atc23/presentation/qiu-haoran},
publisher = {USENIX Association},
month = jul
}

@inproceedings {FIRM,
author = {Haoran Qiu and Subho S. Banerjee and Saurabh Jha and Zbigniew T. Kalbarczyk and Ravishankar K. Iyer},
title = {{FIRM}: An Intelligent Fine-grained Resource Management Framework for {SLO-Oriented} Microservices},
booktitle = {14th USENIX Symposium on Operating Systems Design and Implementation (OSDI 20)},
year = {2020},
isbn = {978-1-939133-19-9},
pages = {805--825},
url = {https://www.usenix.org/conference/osdi20/presentation/qiu},
publisher = {USENIX Association},
month = nov
}

@ARTICLE{al2017elasticity,  author={Al-Dhuraibi, Yahya and Paraiso, Fawaz and Djarallah, Nabil and Merle, Philippe},  journal={IEEE Transactions on Services Computing},   title={Elasticity in Cloud Computing: State of the Art and Research Challenges},   year={2018},  volume={11},  number={2},  pages={430-447},  doi={10.1109/TSC.2017.2711009}}

@INPROCEEDINGS{DeepScaler,
  author={Meng, Chunyang and Song, Shijie and Tong, Haogang and Pan, Maolin and Yu, Yang},
  booktitle={2023 38th IEEE/ACM International Conference on Automated Software Engineering (ASE)}, 
  title={DeepScaler: Holistic Autoscaling for Microservices Based on Spatiotemporal GNN with Adaptive Graph Learning}, 
  year={2023},
  volume={},
  number={},
  pages={53-65},
  doi={10.1109/ASE56229.2023.00038}}

@ARTICLE{ARAScaler,
  author={Jeong, Byeonghui and Jeong, Young-Sik},
  journal={IEEE Transactions on Services Computing}, 
  title={ARAScaler: Adaptive Resource Autoscaling Scheme Using ETimeMixer for Efficient Cloud-Native Computing}, 
  year={2025},
  volume={18},
  number={1},
  pages={72-84},
  doi={10.1109/TSC.2024.3522815}}

@ARTICLE{li2022serverless,
  author={Li, Yongkang and Lin, Yanying and Wang, Yang and Ye, Kejiang and Xu, Chengzhong},
  journal={IEEE Transactions on Services Computing}, 
  title={Serverless Computing: State-of-the-Art, Challenges and Opportunities}, 
  year={2023},
  volume={16},
  number={2},
  pages={1522-1539},
  doi={10.1109/TSC.2022.3166553}}

@ARTICLE{9744560,
  author={Rossi, Fabiana and Cardellini, Valeria and Presti, Francesco Lo and Nardelli, Matteo},
  journal={IEEE Transactions on Cloud Computing}, 
  title={Dynamic Multi-Metric Thresholds for Scaling Applications Using Reinforcement Learning}, 
  year={2023},
  volume={11},
  number={2},
  pages={1807-1821},
  doi={10.1109/TCC.2022.3163357}}

@online{alibabacloud,
  key =          {Alibaba Cloud},
  title="{What is Auto Scaling?}",
  url="https://www.alibabacloud.com/help/en/auto-scaling/product-overview/what-is-auto-scaling",
  lastaccessed = "Feb 10, 2025",
}

@online{wikimedia_rest_api,
  key =          {Wikimedia REST API},
  title="{What is Auto Scaling?}",
  url="https://wikimedia.org/api/rest_v1/",
  lastaccessed = "Feb 10, 2025",
}

@ARTICLE{chen2016self,
  author={Chen, Tao and Bahsoon, Rami},
  journal={IEEE Transactions on Software Engineering}, 
  title={Self-Adaptive and Online QoS Modeling for Cloud-Based Software Services}, 
  year={2017},
  volume={43},
  number={5},
  pages={453-475},
  doi={10.1109/TSE.2016.2608826}}

@article{buyya2018manifesto,
  author = {Buyya, Rajkumar and Srirama, Satish Narayana and Casale, Giuliano and Calheiros, Rodrigo and Simmhan, Yogesh and Varghese, Blesson and Gelenbe, Erol and Javadi, Bahman and Vaquero, Luis Miguel and Netto, Marco A. S. and Toosi, Adel Nadjaran and Rodriguez, Maria Alejandra and Llorente, Ignacio M. and Vimercati, Sabrina De Capitani Di and Samarati, Pierangela and Milojicic, Dejan and Varela, Carlos and Bahsoon, Rami and Assuncao, Marcos Dias De and Rana, Omer and Zhou, Wanlei and Jin, Hai and Gentzsch, Wolfgang and Zomaya, Albert Y. and Shen, Haiying},
  title = {A Manifesto for Future Generation Cloud Computing: Research Directions for the Next Decade},
  year = {2018},
  issue_date = {September 2019},
  publisher = {Association for Computing Machinery},
  address = {New York, NY, USA},
  volume = {51},
  number = {5},
  issn = {0360-0300},
  url = {https://doi.org/10.1145/3241737},
  doi = {10.1145/3241737},
  journal = {ACM Comput. Surv.},
  month = {nov},
  articleno = {105},
  numpages = {38}
  }

@article{qu2018auto,
author = {Qu, Chenhao and Calheiros, Rodrigo N. and Buyya, Rajkumar},
title = {Auto-Scaling Web Applications in Clouds: A Taxonomy and Survey},
year = {2018},
issue_date = {July 2019},
publisher = {Association for Computing Machinery},
address = {New York, NY, USA},
volume = {51},
number = {4},
issn = {0360-0300},
url = {https://doi.org/10.1145/3148149},
doi = {10.1145/3148149},
journal = {ACM Comput. Surv.},
month = {jul},
articleno = {73},
numpages = {33}
}

@inproceedings{10.1145/3542929.3563469,
author = {Wang, Ziliang and Zhu, Shiyi and Li, Jianguo and Jiang, Wei and Ramakrishnan, K. K. and Zheng, Yangfei and Yan, Meng and Zhang, Xiaohong and Liu, Alex X.},
title = {DeepScaling: microservices autoscaling for stable CPU utilization in large scale cloud systems},
year = {2022},
isbn = {9781450394147},
publisher = {Association for Computing Machinery},
address = {New York, NY, USA},
url = {https://doi.org/10.1145/3542929.3563469},
doi = {10.1145/3542929.3563469},
booktitle = {Proceedings of the 13th Symposium on Cloud Computing},
pages = {16–30},
numpages = {15},
location = {San Francisco, California},
series = {SoCC '22}
}

@ARTICLE{abdullah2020burst, 
  author={Abdullah, Muhammad and Iqbal, Waheed and Berral, Josep Lluis and Polo, Jorda and Carrera, David},  
  journal={IEEE Transactions on Services Computing},   title={Burst-Aware Predictive Autoscaling for Containerized Microservices},   
  year={2022},  
  volume={15},  
  number={3},  
  pages={1448-1460},  
  doi={10.1109/TSC.2020.2995937}
}

@article{yin2014system,
title = {System resource utilization analysis and prediction for cloud based applications under bursty workloads},
journal = {Information Sciences},
volume = {279},
pages = {338-357},
year = {2014},
issn = {0020-0255},
doi = {https://doi.org/10.1016/j.ins.2014.03.123},
author = {Jianwei Yin and Xingjian Lu and Hanwei Chen and Xinkui Zhao and Neal N. Xiong}
}

@InProceedings{trihinas2017improving,
author="Trihinas, Demetris
and Georgiou, Zacharias
and Pallis, George
and Dikaiakos, Marios D.",
editor="Alistarh, Dan
and Delis, Alex
and Pallis, George",
title="Improving Rule-Based Elasticity Control by Adapting the Sensitivity of the Auto-Scaling Decision Timeframe",
booktitle="Algorithmic Aspects of Cloud Computing",
year="2018",
publisher="Springer International Publishing",
address="Cham",
pages="123--137",
isbn="978-3-319-74875-7"
}

@ARTICLE{schulman2017proximal,
       author = {{Schulman}, John and {Wolski}, Filip and {Dhariwal}, Prafulla and {Radford}, Alec and {Klimov}, Oleg},
        title = "{Proximal Policy Optimization Algorithms}",
      journal = {arXiv e-prints},
         year = 2017,
        month = jul,
          eid = {arXiv:1707.06347},
        pages = {arXiv:1707.06347},
archivePrefix = {arXiv},
       eprint = {1707.06347},
 primaryClass = {cs.LG},
       adsurl = {https://ui.adsabs.harvard.edu/abs/2017arXiv170706347S}
}

@ARTICLE{bauer2018chameleon,
  author={Bauer, André and Herbst, Nikolas and Spinner, Simon and Ali-Eldin, Ahmed and Kounev, Samuel},
  journal={IEEE Transactions on Parallel and Distributed Systems}, 
  title={Chameleon: A Hybrid, Proactive Auto-Scaling Mechanism on a Level-Playing Field}, 
  year={2019},
  volume={30},
  number={4},
  pages={800-813},
  doi={10.1109/TPDS.2018.2870389}}

@ARTICLE{yu2020microscaler,
  author={Yu, Guangba and Chen, Pengfei and Zheng, Zibin},
  journal={IEEE Transactions on Cloud Computing}, 
  title={Microscaler: Cost-Effective Scaling for Microservice Applications in the Cloud With an Online Learning Approach}, 
  year={2022},
  volume={10},
  number={2},
  pages={1100-1116},
  doi={10.1109/TCC.2020.2985352}}

@article{gari2021reinforcement,
title = {Reinforcement learning-based application Autoscaling in the Cloud: A survey},
journal = {Engineering Applications of Artificial Intelligence},
volume = {102},
pages = {104288},
year = {2021},
issn = {0952-1976},
doi = {https://doi.org/10.1016/j.engappai.2021.104288},
url = {https://www.sciencedirect.com/science/article/pii/S0952197621001354},
author = {Yisel Garí and David A. Monge and Elina Pacini and Cristian Mateos and Carlos {García Garino}}
}

@INPROCEEDINGS{schuler2020ai,  author={Schuler, Lucia and Jamil, Somaya and Kühl, Niklas},  booktitle={2021 IEEE/ACM 21st International Symposium on Cluster, Cloud and Internet Computing (CCGrid)},   title={AI-based Resource Allocation: Reinforcement Learning for Adaptive Auto-scaling in Serverless Environments},   year={2021},  volume={},  number={},  pages={804-811},  doi={10.1109/CCGrid51090.2021.00098}}

@article{yan2021hansel,
title = {HANSEL: Adaptive horizontal scaling of microservices using Bi-LSTM},
journal = {Applied Soft Computing},
volume = {105},
pages = {107216},
year = {2021},
issn = {1568-4946},
doi = {https://doi.org/10.1016/j.asoc.2021.107216},
author = {Ming Yan and XiaoMeng Liang and ZhiHui Lu and Jie Wu and Wei Zhang}
}

@INPROCEEDINGS{zhang2020sarsa,  author={Zhang, Shubo and Wu, Tianyang and Pan, Maolin and Zhang, Chaomeng and Yu, Yang},  booktitle={2020 IEEE International Conference on Web Services (ICWS)},   title={A-SARSA: A Predictive Container Auto-Scaling Algorithm Based on Reinforcement Learning},   year={2020},  volume={},  number={},  pages={489-497},  doi={10.1109/ICWS49710.2020.00072}}

@INPROCEEDINGS{ali2014measuring,  author={Ali-Eldin, Ahmed and Seleznjev, Oleg and Sjöstedt-de Luna, Sara and Tordsson, Johan and Elmroth, Erik},  booktitle={2014 IEEE/ACM 7th International Conference on Utility and Cloud Computing},   title={Measuring Cloud Workload Burstiness},   year={2014},  volume={},  number={},  pages={566-572},  doi={10.1109/UCC.2014.87}}

@article{zhou2021informer, title={Informer: Beyond Efficient Transformer for Long Sequence Time-Series Forecasting}, volume={35}, url={https://ojs.aaai.org/index.php/AAAI/article/view/17325}, abstractNote={Many real-world applications require the prediction of long sequence time-series, such as electricity consumption planning. Long sequence time-series forecasting (LSTF) demands a high prediction capacity of the model, which is the ability to capture precise long-range dependency coupling between output and input efficiently. Recent studies have shown the potential of Transformer to increase the prediction capacity. However, there are several severe issues with Transformer that prevent it from being directly applicable to LSTF, including quadratic time complexity, high memory usage, and inherent limitation of the encoder-decoder architecture. To address these issues, we design an efficient transformer-based model for LSTF, named Informer, with three distinctive characteristics: (i) a ProbSparse self-attention mechanism, which achieves O(L log L) in time complexity and memory usage, and has comparable performance on sequences’ dependency alignment. (ii) the self-attention distilling highlights dominating attention by halving cascading layer input, and efficiently handles extreme long input sequences. (iii) the generative style decoder, while conceptually simple, predicts the long time-series sequences at one forward operation rather than a step-by-step way, which drastically improves the inference speed of long-sequence predictions. Extensive experiments on four large-scale datasets demonstrate that Informer significantly outperforms existing methods and provides a new solution to the LSTF problem.}, number={12}, journal={Proceedings of the AAAI Conference on Artificial Intelligence}, author={Zhou, Haoyi and Zhang, Shanghang and Peng, Jieqi and Zhang, Shuai and Li, Jianxin and Xiong, Hui and Zhang, Wancai}, year={2021}, month={May}, pages={11106-11115} }

@article{zhong2022machine,
author = {Zhong, Zhiheng and Xu, Minxian and Rodriguez, Maria Alejandra and Xu, Chengzhong and Buyya, Rajkumar},
title = {Machine Learning-Based Orchestration of Containers: A Taxonomy and Future Directions},
year = {2022},
publisher = {Association for Computing Machinery},
address = {New York, NY, USA},
issn = {0360-0300},
url = {https://doi.org/10.1145/3510415},
doi = {10.1145/3510415},
note = {Just Accepted},
journal = {ACM Comput. Surv.},
month = {jan}
}

@ARTICLE{10433234,
  author={Deng, Shuiguang and Zhao, Hailiang and Huang, Binbin and Zhang, Cheng and Chen, Feiyi and Deng, Yinuo and Yin, Jianwei and Dustdar, Schahram and Zomaya, Albert Y.},
  journal={Proceedings of the IEEE}, 
  title={Cloud-Native Computing: A Survey From the Perspective of Services}, 
  year={2024},
  volume={112},
  number={1},
  pages={12-46},
  doi={10.1109/JPROC.2024.3353855}}

@INPROCEEDINGS{ding2021copa,  author={Ding, Zhijun and Huang, Qichen},  booktitle={2021 IEEE International Conference on Web Services (ICWS)},   title={COPA: A Combined Autoscaling Method for Kubernetes},   year={2021},  volume={},  number={},  pages={416-425},  doi={10.1109/ICWS53863.2021.00061}}

@inproceedings{vaswani2017attention,
 author = {Vaswani, Ashish and Shazeer, Noam and Parmar, Niki and Uszkoreit, Jakob and Jones, Llion and Gomez, Aidan N and Kaiser, \L ukasz and Polosukhin, Illia},
 booktitle = {Advances in Neural Information Processing Systems},
 editor = {I. Guyon and U. Von Luxburg and S. Bengio and H. Wallach and R. Fergus and S. Vishwanathan and R. Garnett},
 pages = {},
 publisher = {Curran Associates, Inc.},
 title = {Attention is All you Need},
 url = {https://proceedings.neurips.cc/paper/2017/file/3f5ee243547dee91fbd053c1c4a845aa-Paper.pdf},
 volume = {30},
 year = {2017}
}

@article{awad2015support,
  title={Support vector regression},
  author={Awad, Mariette and Khanna, Rahul and Awad, Mariette and Khanna, Rahul},
  journal={Efficient learning machines: Theories, concepts, and applications for engineers and system designers},
  pages={67--80},
  year={2015},
  publisher={Springer}
}

@article{bollerslev1986generalized,
title = {Generalized autoregressive conditional heteroskedasticity},
journal = {Journal of Econometrics},
volume = {31},
number = {3},
pages = {307-327},
year = {1986},
issn = {0304-4076},
doi = {https://doi.org/10.1016/0304-4076(86)90063-1},
url = {https://www.sciencedirect.com/science/article/pii/0304407686900631},
author = {Tim Bollerslev}
}

@article{masdari2020survey,
  title={A survey and classification of the workload forecasting methods in cloud computing},
  author={Masdari, Mohammad and Khoshnevis, Afsane},
  journal={Cluster Computing},
  volume={23},
  number={4},
  pages={2399--2424},
  year={2020},
  publisher={Springer}
}

@article{calzarossa2016workload,
  title={Workload characterization: A survey revisited},
  author={Calzarossa, Maria Carla and Massari, Luisa and Tessera, Daniele},
  journal={ACM Computing Surveys (CSUR)},
  volume={48},
  number={3},
  pages={1--43},
  year={2016},
  publisher={ACM New York, NY, USA}
}

@article{chen2022resource,
  title={Resource Allocation with Workload-Time Windows for Cloud-Based Software Services: A Deep Reinforcement Learning Approach},
  author={Chen, Xing and Yang, Lijian and Chen, Zheyi and Min, Geyong and Zheng, Xianghan and Rong, Chunming},
  journal={IEEE Transactions on Cloud Computing},
  year={2022},
  publisher={IEEE}
}

@article{golshani2021proactive,
  title={Proactive auto-scaling for cloud environments using temporal convolutional neural networks},
  author={Golshani, Ehsan and Ashtiani, Mehrdad},
  journal={Journal of Parallel and Distributed Computing},
  volume={154},
  pages={119--141},
  year={2021},
  publisher={Elsevier}
}

@article{imdoukh2020machine,
  title={Machine learning-based auto-scaling for containerized applications},
  author={Imdoukh, Mahmoud and Ahmad, Imtiaz and Alfailakawi, Mohammad Gh},
  journal={Neural Computing and Applications},
  volume={32},
  number={13},
  pages={9745--9760},
  year={2020},
  publisher={Springer}
}

@article{ramperez2021flas,
  title={Flas: A combination of proactive and reactive auto-scaling architecture for distributed services},
  author={Ramp{\'e}rez, V{\'\i}ctor and Soriano, Javier and Lizcano, David and Lara, Juan A},
  journal={Future Generation Computer Systems},
  volume={118},
  pages={56--72},
  year={2021},
  publisher={Elsevier}
}

@article{tadakamalla2020autonomic,
  title={Autonomic elasticity control for multi-server queues under generic workload surges in cloud environments},
  author={Tadakamalla, Venkat and Menasce, Daniel},
  journal={IEEE Transactions on Cloud Computing},
  year={2020},
  publisher={IEEE}
}

@inproceedings{vlachos2004identifying,
author = {Vlachos, Michail and Meek, Christopher and Vagena, Zografoula and Gunopulos, Dimitrios},
title = {Identifying Similarities, Periodicities and Bursts for Online Search Queries},
year = {2004},
isbn = {1581138598},
publisher = {Association for Computing Machinery},
address = {New York, NY, USA},
url = {https://doi.org/10.1145/1007568.1007586},
doi = {10.1145/1007568.1007586},
booktitle = {Proceedings of the 2004 ACM SIGMOD International Conference on Management of Data},
pages = {131–142},
numpages = {12},
location = {Paris, France},
series = {SIGMOD '04}
}

@article{verma2021auto,
  title={Auto-scaling techniques for IoT-based cloud applications: a review},
  author={Verma, Shveta and Bala, Anju},
  journal={Cluster Computing},
  volume={24},
  number={3},
  pages={2425--2459},
  year={2021},
  publisher={Springer}
}

@article{TARI2024100650,
title = {Auto-scaling mechanisms in serverless computing: A comprehensive review},
journal = {Computer Science Review},
volume = {53},
pages = {100650},
year = {2024},
issn = {1574-0137},
doi = {https://doi.org/10.1016/j.cosrev.2024.100650},
author = {Mohammad Tari and Mostafa Ghobaei-Arani and Jafar Pouramini and Mohsen Ghorbian}
}

@ARTICLE{10419899,
  author={Quattrocchi, Giovanni and Incerto, Emilio and Pinciroli, Riccardo and Trubiani, Catia and Baresi, Luciano},
  journal={IEEE Transactions on Services Computing}, 
  title={Autoscaling Solutions for Cloud Applications Under Dynamic Workloads}, 
  year={2024},
  volume={17},
  number={3},
  pages={804-820},
  doi={10.1109/TSC.2024.3354062}}

@ARTICLE{9740415,
  author={He, Xiang and Tu, Zhiying and Wagner, Markus and Xu, Xiaofei and Wang, Zhongjie},
  journal={IEEE Transactions on Cloud Computing}, 
  title={Online Deployment Algorithms for Microservice Systems With Complex Dependencies}, 
  year={2023},
  volume={11},
  number={2},
  pages={1746-1763},
  doi={10.1109/TCC.2022.3161684}}

@inproceedings{
wang2024timemixer,
title={TimeMixer: Decomposable Multiscale Mixing for Time Series Forecasting},
author={Shiyu Wang and Haixu Wu and Xiaoming Shi and Tengge Hu and Huakun Luo and Lintao Ma and James Y. Zhang and JUN ZHOU},
booktitle={The Twelfth International Conference on Learning Representations},
year={2024},
url={https://openreview.net/forum?id=7oLshfEIC2}
}

@INPROCEEDINGS{9763776,
  author={Li, Ye and Zhang, Haitao and Tian, Wei and Ma, Huadong},
  booktitle={2021 IEEE 27th International Conference on Parallel and Distributed Systems (ICPADS)}, 
  title={Joint Optimization of Auto-Scaling and Adaptive Service Placement in Edge Computing}, 
  year={2021},
  volume={},
  number={},
  pages={923-930},
  doi={10.1109/ICPADS53394.2021.00121}}

@ARTICLE{cheng2023proscale,
  author={Cheng, Ke and Zhang, Sheng and Tu, Chenghong and Shi, Xiaohang and Yin, Zhaoheng and Lu, Sanglu and Liang, Yu and Gu, Qing},
  journal={IEEE Transactions on Parallel and Distributed Systems}, 
  title={ProScale: Proactive Autoscaling for Microservice With Time-Varying Workload At the Edge}, 
  year={2023},
  volume={},
  number={},
  pages={1-18},
  doi={10.1109/TPDS.2023.3238429}}

@INPROCEEDINGS{qian2022robustscaler,
  author={Qian, Huajie and Wen, Qingsong and Sun, Liang and Gu, Jing and Niu, Qiulin and Tang, Zhimin},
  booktitle={2022 IEEE 38th International Conference on Data Engineering (ICDE)}, 
  title={RobustScaler: QoS-Aware Autoscaling for Complex Workloads}, 
  year={2022},
  volume={},
  number={},
  pages={2762-2775},
  doi={10.1109/ICDE53745.2022.00252}}

@article{zargarazad2023auto,
  title={An auto-scaling approach for microservices in cloud computing environments},
  author={ZargarAzad, Matineh and Ashtiani, Mehrdad},
  journal={Journal of Grid Computing},
  volume={21},
  number={4},
  pages={73},
  year={2023},
  publisher={Springer}
}

@INPROCEEDINGS{9671185,
  author={Perera, H. C. S. and De Silva, T. S. D. and Wasala, W. M. D. C. and Rajapakshe, R. M. P. R. L. and Kodagoda, N. and Samaratunge, Udara Srimath S. and Jayanandana, H. H. N. C.},
  booktitle={2021 3rd International Conference on Advancements in Computing (ICAC)}, 
  title={Database Scaling on Kubernetes}, 
  year={2021},
  volume={},
  number={},
  pages={258-263},
  doi={10.1109/ICAC54203.2021.9671185}}

@ARTICLE{9328525,
  author={Toka, László and Dobreff, Gergely and Fodor, Balázs and Sonkoly, Balázs},
  journal={IEEE Transactions on Network and Service Management}, 
  title={Machine Learning-Based Scaling Management for Kubernetes Edge Clusters}, 
  year={2021},
  volume={18},
  number={1},
  pages={958-972},
  doi={10.1109/TNSM.2021.3052837}}

@electronic{DFT,
  title="{DFT: A function service that comput DFT and its inverse}",
  url="https://github.com/wtysos11/DFTService",
  year = 2024,
  note = "Accessed: Aug 10, 2024",
}

@electronic{sockshop,
  title="{Sock Shop: A Microservice Demo Application}",
  url="https://github.com/microservices-demo/microservices-demo",
  year = 2024,
  note = "Accessed: Aug 10, 2024",
}

@electronic{bookinfo,
  title="{Bookinfo: A microservice benchmark provided by Istio}",
  url="https://istio.io/docs/examples/bookinfo/",
  year = 2024,
  note = "Accessed: Aug 10, 2024",
}

@electronic{trainticket,
  title="{Train Ticket: A Benchmark Microservice System}",
  url="https://github.com/FudanSELab/train-ticket",
  year = 2024,
  note = "Accessed: Aug 10, 2024",
}

@electronic{boutique,
  title="{Online Boutique: A microservice benchmark used by Google Cloud Platform}",
  url="https://github.com/GoogleCloudPlatform/microservices-demo",
  year = 2024,
  note = "Accessed: Aug 10, 2024",
}

@online{grafana_k6,
  title="{Grafana K6: a modern load-testing tool}",
  url="https://github.com/grafana/k6",
  lastaccessed = "Feb 10, 2025",
}

@online{svr_impl,
  title="{The implementation of class sklearn.svm.SVR}",
  url="https://scikit-learn.org/stable/modules/generated/sklearn.svm.SVR.html",
  lastaccessed = "Feb 10, 2025",
}

@online{informer_impl,
  title="{The origin Pytorch implementation of Informer}",
  url="https://github.com/zhouhaoyi/Informer2020",
  lastaccessed = "Feb 10, 2025",
}

@online{ar_impl,
  title="{Statsmodels: a Python package that provides a complement to scipy for statistical computations}",
  url="https://github.com/statsmodels/statsmodels",
  lastaccessed = "Feb 10, 2025",
}

@online{boostrapping_impl,
  title="{Bootstrapped - confidence intervals made easy}",
  url="https://github.com/facebookarchive/bootstrapped",
  lastaccessed = "Feb 10, 2025",
}

@online{ElegantRL,
  title="{ElegantRL: Massively Parallel Deep Reinforcement Learning}",
  url="https://github.com/AI4Finance-Foundation/ElegantRL",
  lastaccessed = "Feb 10, 2025",
}

@online{istio,
  key =          {Istio},
  year = 2022,
  title="{An open source service mesh that layers transparently onto existing distributed applications}",
  url="https://istio.io/",
  lastaccessed = "Feb 10, 2025",
}

@online{k8sapi,
  key =          {Kubernetes Clients},
  year  =        2022,
  title =        "Python client for the kubernetes API",
  url =          "https://github.com/kubernetes-client/python",
  lastaccessed = "Feb 10, 2025",
}

@online{k8sHPA,
  key =          {HPA},
  year  =        2021,
  title =        "Horizontal Pod Autoscaling | Kubernetes",
  url =          "https://kubernetes.io/docs/tasks/run-application/horizontal-pod-autoscale/",
  lastaccessed = "Feb 10, 2025",
}

@online{Google,
  key =          {Dan Paik},
  year  =        2016,
  title =        "Adapt or Die: A microservices story at Google",
  url =          "https://www.slideshare.net/apigee/adapt-or-diea-microservices-story-at-google",
  lastaccessed = "Oct 10, 2023",
}

\vspace{-3em}
\begin{IEEEbiography}[{\includegraphics[width=1in,height=1.25in,clip,keepaspectratio]{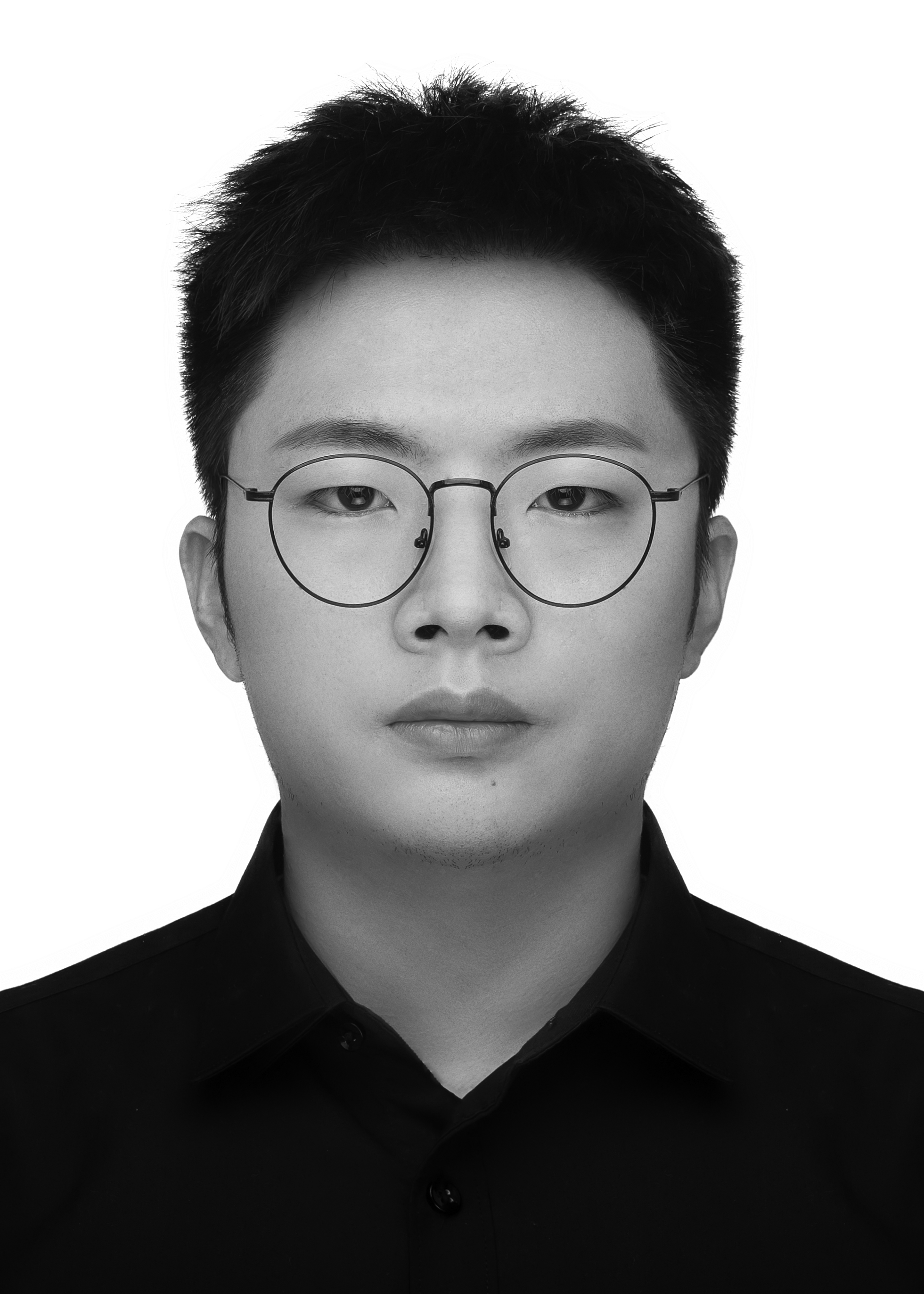}}]{Chunyang Meng}
received his Ph.D. degree in computer science from the School of Computer Science, Sun Yat-sen University, China, in 2024. He is currently a postdoctoral researcher with the School of Computer Science, Chongqing University of Posts and Telecommunications, China. His current research areas include services computing and AI driven operations. He was a recipient of the ACM SIGSOFT Distinguished Paper Award at ASE 2023 and the Best Student Paper Award at ICWS 2022.
\end{IEEEbiography}

\vspace{-3em}
\begin{IEEEbiography}[{\includegraphics[width=1in,height=1.25in,clip,keepaspectratio]{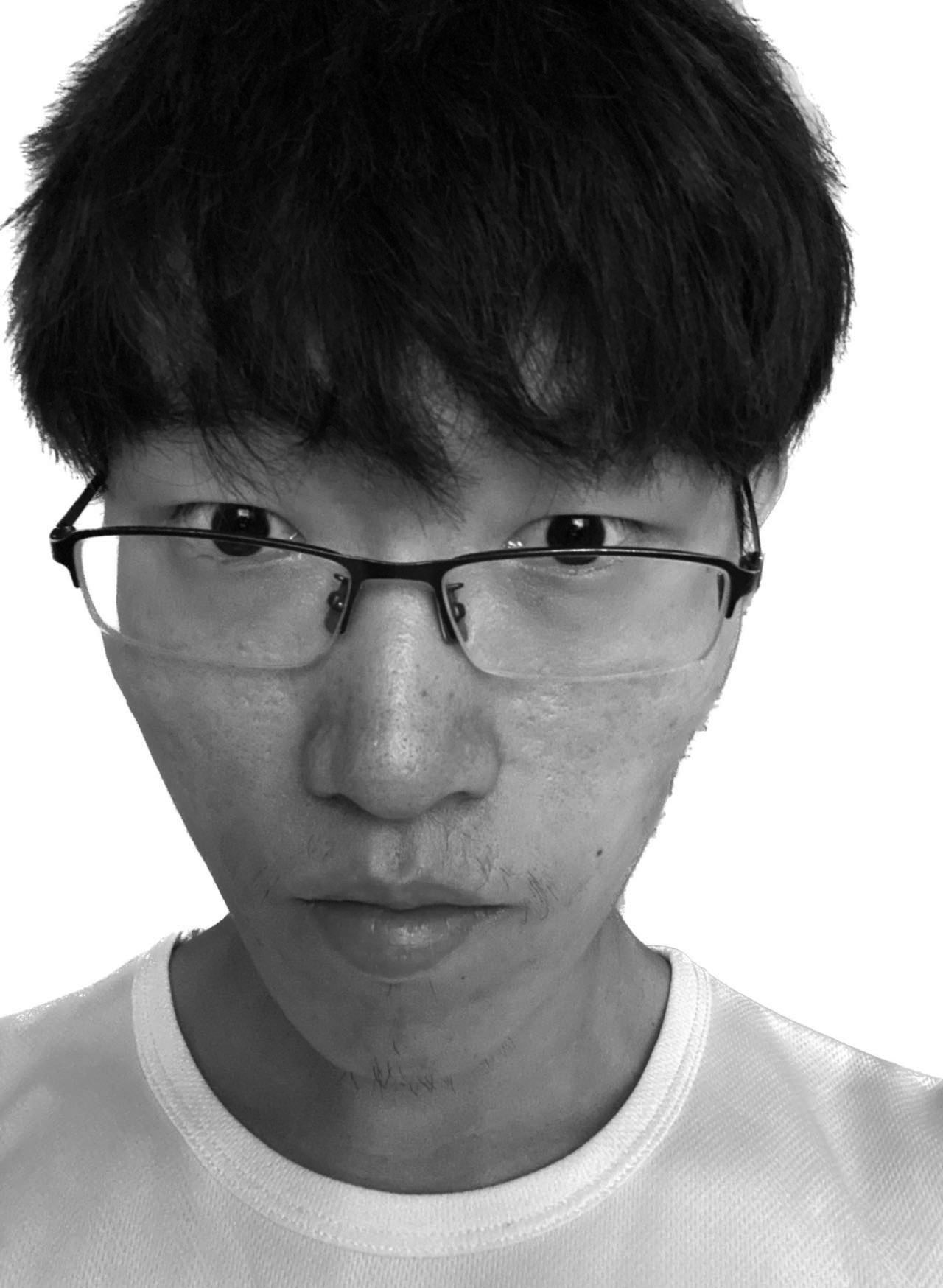}}]{Haogang Tong} was born in 1995. He obtained his master's degree from Beijing University of Posts and Telecommunications in 2020. Currently, he is pursuing a Ph.D. degree at Sun Yat-sen University. He was a recipient of the ACM SIGSOFT Distinguished Paper Award. His research interests include microservices scheduling, edge computing, and the Internet of Things (IoT).
\end{IEEEbiography}

\vspace{-3em}
\begin{IEEEbiography}[{\includegraphics[width=1in,height=1.25in,clip,keepaspectratio]{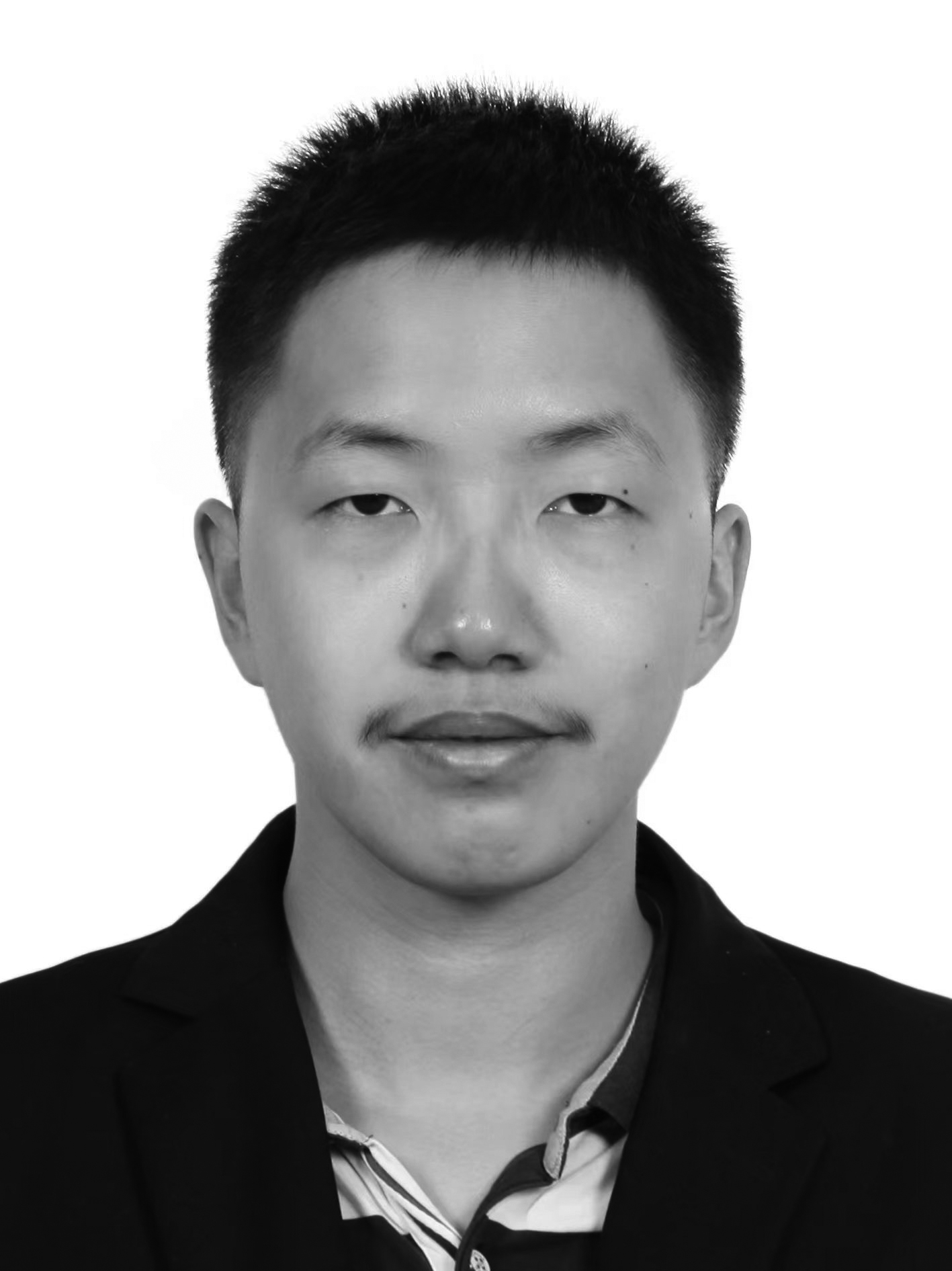}}]{Tianyang Wu} received the bachelor's and master's degrees in software engineering from the Sun Yat-sen University, Guangzhou, China, in 2020, and 2022 respectively. He is now working for Tencent Co. Ltd, China. His research interests include service computing, cloud computing and AI driven operations.
\end{IEEEbiography}

\vspace{-3em}
\begin{IEEEbiography}[{\includegraphics[width=1in,height=1.25in,clip,keepaspectratio]{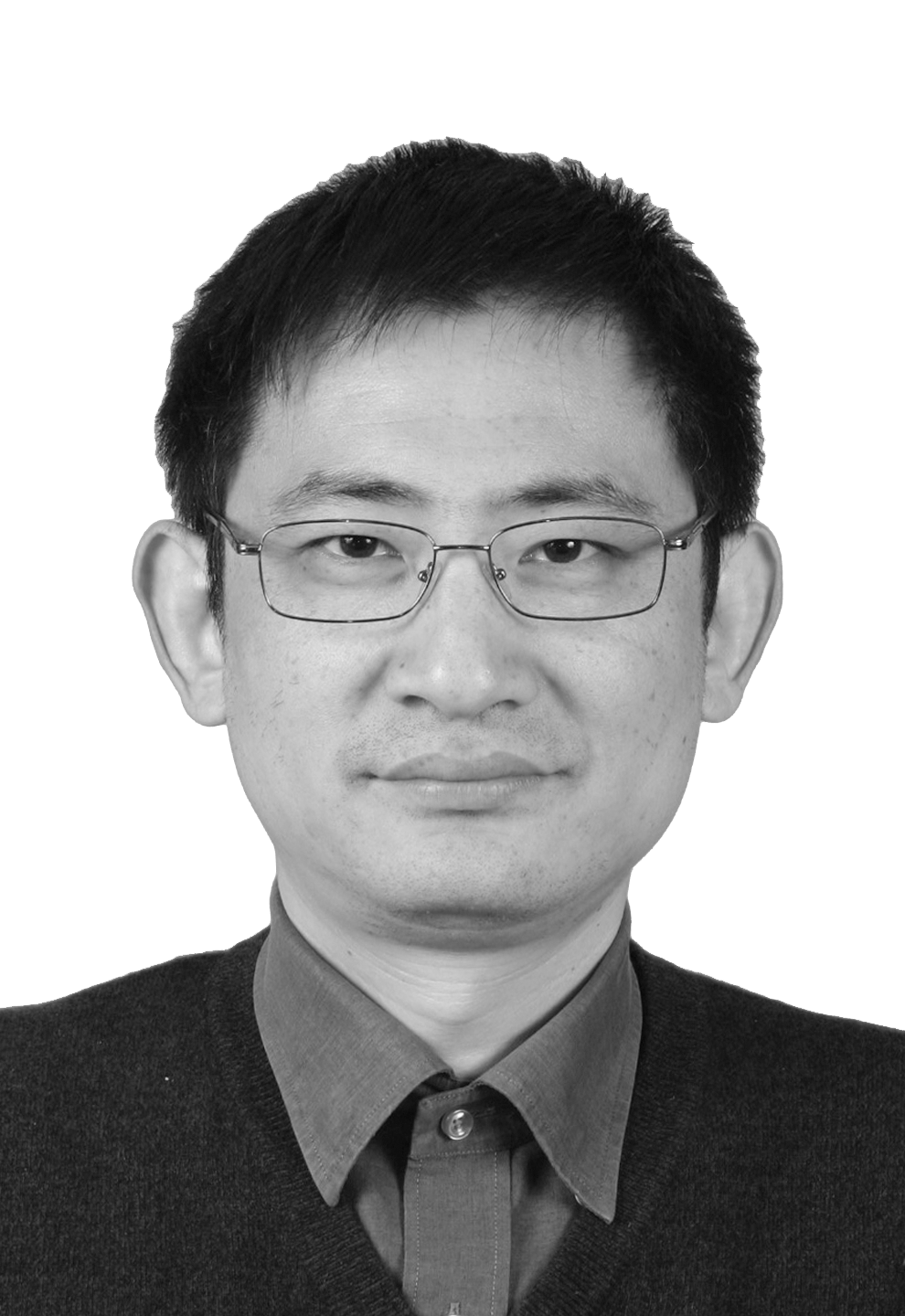}}]{Maolin Pan} received the B.S. degree from the Huazhong University of Science and Technology (HUST), Wuhan, China, in 1988, and the M.S. degree from HUST, in 1991, and the Ph.D. degree from Sun Yat-sen University (SYSU), Guangzhou, China, in 2017. He is currently a lecturer with the School of Software Engineering, SYSU. His research interests include BPM and workflow technical, cloud native computing.
\end{IEEEbiography}

\vspace{-3em}
\begin{IEEEbiography}[{\includegraphics[width=1in,height=1.25in,clip,keepaspectratio]{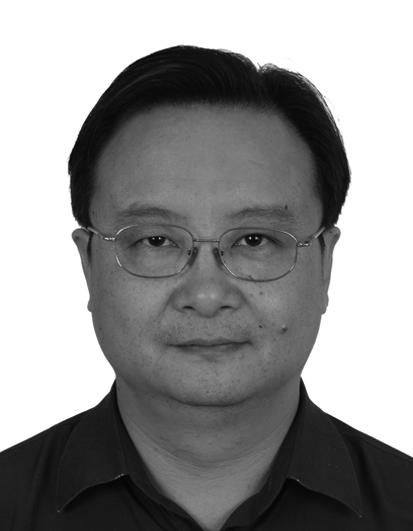}}]{Yang Yu} received the Ph.D. degree in computer science from Sun Yat-sen University (SYSU), Guangzhou, China. 
He is currently a full professor with the School of Software Engineering, SYSU.
He has published more than 90 papers and has been a reviewer for several prestigious international conferences and journals. He is a Distinguished Member of CCF and a member of ACM. His research interests include workflow/BPM, service computing, cloud computing, and software engineering.
\end{IEEEbiography}

\vspace{-3em}
\begin{IEEEbiography}[{\includegraphics[width=1in,height=1.25in,clip,keepaspectratio]{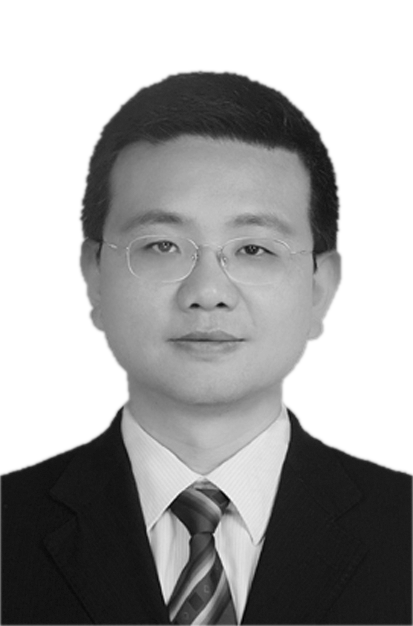}}]{Yi Jiang} received the Ph.D. degree in computer science from the School of Computer Science and Engineering, University of Electronic Science and Technology, China, in 2015. He is currently a professor with the School of Computer Science and Technology, Chongqing University of Posts and Telecommunications, Chongqing, China.
His current research interests include computer architecture, software engineering, big data, and network security.
\end{IEEEbiography}

\end{document}